\documentclass[12pt]{article}

\usepackage{graphics}
\usepackage{amsmath}
\def\ni{\noindent}
\def\ph{{\phantom{...}}}
\def\={\phantom{..} = \phantom{..}}

\def\+{\phantom{..} + \phantom{..}}
\def\>{\phantom{..} > \phantom{..}}
\def\<{\phantom{..} < \phantom{..}}
\def\-{\phantom{..} - \phantom{..}}
\def\Geq{\phantom{..} \geq  \phantom{..}}
\def\Leq{\phantom{..} \leq  \phantom{..}}

\def\bq{\begin{quote}}
\def\eq{\end{quote}}
\def\be{\begin{equation}}
\def\ee{\end{equation}}
\def\bar{\begin{eqnarray}}
\def\ear{\end{eqnarray}}
\def\no{\nonumber}

\def\Sch{Schr{\"o}dinger}

\def\Schists{Schr{\"o}dingerists}

\def\Re{\hbox{Re}}

\def\cH{{\bf{H}}}

\def\sjN{\sum_{j=1}^N}

\def\sjkN{\sum_{j,k=1}^N}
\def\skN{\sum_{k=1}^N}

\def\Re{\hbox{Re}}

\def\srmo{{\sqrt{ - 1}}}
\def\half{{\frac{1}{2}}}

\def\diags{\hbox{diag}\,(s)}
\def\diagss{\hbox{diag}\,(s^2)}

\def\cE{{\cal E}}
\def\un{\underline{n}}
\def\um{\underline{m}}
\def\Ntup{x_1,x_2,...,x_N}
\def\ootm{\left(\,\frac{1}{2\,m}\,\right)}  
  
\def\ootri{- \ootm\,\triangle}

\def\ovx{\overline{x}}
\def\pdxk{\prod\,dx_k}

\def\ipdzk{\int\,\prod\,dz_k}
\def\LtRN{L^2(R^N)}
\def\sdic{sensitive dependence on initial conditions}
\def\cEz{\cE_{\underline{0}}}
\def\diag{\hbox{diag}}
\def\cC{{\cal C}}
\def\cCe{{\cal C}e}
\def\cCo{{\cal C}o}
\def\Runum{R_{\un,\um}}
\def\aun{a_{\un}}
\def\aum{a_{\um}}
\def\sunC{\sum_{\un \in \cC}}
\def\sumC{\sum_{\um \in \cC}}
\def\Lun{\Lambda_{\un}}
\def\cR{{\cal R}}
\def\cala{{\cal A}}

\title{\bf Chaos in a Nonlinear Wavefunction Model:\\[0.5in]
 An Alternative to Born's Probability Hypothesis\\[3in]}

\author{W. David Wick\footnote{email: wdavid.wick@gmail.com}}

\begin{document}
\maketitle
\pagebreak

\section*{Abstract}
In a prior paper, the author described an instability in a nonlinear wavefunction model.
Proposed in connection with the Measurement Problem, 
the model contained an external potential 
creating a ``classical'' instability.
However, it is interesting to ask whether such models  
possess an intrinsic randomness---even ``chaos"---independent of external potentials. 
In this work, 
I investigate the criterion analytically and simulate from a small (``3 qubit")
model, demonstrating that the Lyapunov exponent---a standard measure of ``chaos"---is positive. 
I also extend the instability criterion to models in the continuum.
These results suggest that the boundary between classical and wavefunction physics may also
constitute the threshold of chaos,
and present an alternative to Max Born's {\em ad hoc} probability
hypothesis: random outcomes in experiments
result not from ``wave-particle duality" or ``the existence of the quantum," but {\em 
from sensitive dependence on initial conditions}, as is common in the other sciences. 
 
\section{Introduction}

In an earlier work, \cite{paperIII}, which I will refer to here as ``paper III", 
I introduced into a finite-dimensional
(``multi-qubit") setting some nonlinear terms which, in a continuum setting, I earlier
proposed (in \cite{paperI}, ``paper I") as a solution to the Measurement Problem.
In that context I studied a certain matrix, a function of the wavefunction, that I wrote as ``$M$''.
This matrix (known variously as the Hessian or Jacobian of the dynamical system) defines 
a linear approximation to the system's motion for a small (infinitesimal) time.
I then noted that, in the nonlinear model, this matrix could have real eigenvalues, but not
in the linear case\footnote{I'm referring here to the real dynamical system (RDS) 
associated with \Sch's complex wavefunction set-up, see next section.
In that context, linear ``Quantum" models have only imaginary eigenvalues; see paper III,
section 2.} (of ordinary ``Quantum Mechanics"\footnote{Although I labelled 
the first series of articles as about ``Non-Linear Quantum Mechanics", I no longer
use the term ``quantum", as I have realized that I do not know what it refers to.}).
I then showed that the condition: $\det\,M < 0$, implied that $M$ had both positive and
negative eigenvalues, meaning that unstable and stable directions met at that 
point (wavefunction). This appearance, I thought, brought up the specter of ``chaos".

Paper III was written after the author had reconsidered his position, exposed in paper II,
\cite{paperII}, that explaining randomness in experimental outcomes\footnote{\Schists\
do not accept Born's probability hypothesis (the modulus-squared wavefunction represents
``the probability of finding the particle" somewhere). For us, $\psi$ is an element
of reality having nothing to do with probability or statistics. Nor do we accept von
Neumann's measurement axioms.} required the introduction
of a universal, random component of wavefunctions. This apostasy resulted from recalling that
high-dimensional nonlinear dynamical systems often exhibit the various phenomena lumped
together in the word ``chaos". Thus I undertook a search for conditions that might imply
such phenomena and stumbled upon the determinant condition.
 
In the author's view at the time, the determinant condition served primarily
to assure that the model was not essentially equivalent to linear theory (which cannot
exhibit chaos). Meanwhile, the author
suggested that the random outcomes (``the needle on the dial went left, OR it went right" with
certain probabilities), was created by including a double-well potential 
and starting the system near the unstable critical point. 
The role of the nonlinear terms might be to render the dynamics
essentially classical, so the system would resemble a pencil balanced momentarily on its tip 
then given a small random push.

Thus the philosophy was: ``The apparatus creates the randomness
as well as the dichotomy." 
Although an appropriate attitude for addressing the Measurement Problem,
deriving---or attributing---randomness to the intervention
of an external potential is not in the spirit of chaos theory.
Could randomness arise entirely from {\em intrinsic instability},
without assistance from external forces?

In that earlier work, I also failed to make an interesting observation: because the determinant
condition 
takes the form of an
inequality,  
every such wavefunction will exist in a neighborhood also satisfying the condition.
The system, started at such a point, will enjoy that condition at least for a short time.
Thus the main issue I pursue in this paper is: what is the consequence for a dynamical
(Hamiltonian) system if it flows through points with linear instabilities for a period?

Now to ``chaos", a misnomer but a usage too popular today to entirely avoid. It appears to
mean two things: (a) sensitive dependence on initial conditions (``the butterfly flapping its 
wings today in Seattle causes a hurricane in Brazil a century later") and (b) 
random-appearing trajectories. Both of these phenomena, however defined, clearly are a matter
of time scales. For instance, 30 years ago a group demonstrated
 ``chaos" in the outer planets of
the solar system,\cite{MandH}, by simulating it (on a fast computer using a symplectic algorithm), and comparing
two trajectories, the second obtained by making a 1.5 cm shift in the center-of-mass of Uranus.
The trajectories diverged on a time-scale of millions of years and on that time-scale
the planets' motion appeared to be somewhat random
(while on a human time-scale,
say of centuries, the motion is completely predictable and regular).

We can be somewhat more precise about the definition of this mysterious ``chaos".
According to the review article of Skokos, \cite{Skokos}\footnote{I derived most of my
understanding of chaos in Hamiltonian systems from this excellent article. I thank
L. De Carlo for suggesting and providing the article.} in 1989 Devaney gave one of the ``most
famous" definitions: chaos in a Hamiltonian system means three things:

\begin{quote}
(1) Sensitive dependence on initial conditions;

(2) Topological transitivity (the system eventually visits every neighborhood in the state space);

(3) dense periodic orbits.
\end{quote} 

Obviously, conditions (2) and (3) are difficult to check analytically and impossible to verify
by simulations. Moreover, part (1) already captures the essence of randomness, when referring
to experiments: the inability, however much we try to control the relevant variables, to
get our experimental set-up to exactly reproduce prior outcomes. I suspect that (2) and (3)
are meant to get at apparent 
randomness in a time-series of observations during a single experimental
``run" (as might be made by galactic neighbors if they swung by our solar system every
few million years). 

Therefore, I will concentrate in this work on establishing that condition (1) for chaos
holds for certain nonlinear wavefunction 
systems in certain situations. There are two approaches that I will
pursue {\em via} simulations. In one, solve (simulate) the Hamiltonian system from two
nearby initial states and observe the deviation over time. In another, solve a larger
system consisting of the RDS derived from the Hamiltonian simultaneously with the derived
JDS, from some initial choice of system state and one tangent vector;
then observe any growth in the norm of that vector. These approaches, and the pitfalls to avoid,
are further explained, and the simulations plotted, in section \ref{simsection}.

Section \ref{algebrasection} introduces the algebraic analysis of the instability criterion.
Section \ref{continuumsection} describes how these results may be generalized to 
continuum models. 
 
There are five Math Appendicies. Math Appendix 1 contains a list of useful identities
concerning matrix algebra and the determinant function. Math Appendix 2 applies some of 
these identities in an analysis of the determinant condition implying instability; 
two theorems and various corollaries are proven.
Math Appendix 3 applies the theorems in the context of the discrete 
(``spin" or ``qubit") model of paper III.
Math Appendix 4 extends the instability criteria to continuum models.
Math Appendix 5 has some details about the numerical methods used for the simulations.

\section{The Model and the Criterion for Instability\label{modelsection}}

The model is a finite-dimensional Hamiltonian system of wavefunctions on configurations
consisting of three ``spins" or qubits: $(\sigma_1,\sigma_2,\sigma_3)$, with $\sigma_i 
= \pm\,1/2$ for $i= 1,2,3$. Thus the state vector, $\psi$, takes on $N = 8$
 complex values, or 16 real values. The Hamiltonian takes the form:

\def\HQM{H_{QM}}

\be
\cH \= <\psi|\,\HQM\,|\psi> \+ w\,\left\{\,<\psi|\,S^2\,|\psi> 
\- <\psi|\,S\,|\psi>^2\,\right\}.\label{Ham1}
\ee

Next, the wavefunction is decomposed into real and imaginary parts as:

\be
\psi_k \= Q_k \+ \srmo\,P_k,
\ee

\ni where `$k$' here indexes a configuration.\footnote{$Q_k$ and $P_k$ do NOT represent
physical positions and momentums, but rather are 
the Q's and P's of a course in classical mechanics.} 
Plugging into (\ref{Ham1}) and writing $s_k$ for the sum 
 $\sigma_1 + \sigma_2 + \sigma_3$, I obtained in that paper with choices: $S = \hbox{diag}\{s_k\}$
and $\HQM = K$, a real, symmetrical matrix:

\bar
\no \cH &\=& (1/2)\,\sum_{j,k}\,P_j\,K_{j,k}\,P_k 
\+ (1/2)\,\sum_{j,k}\,Q_j\,K_{j,k}\,Q_k\\
\no &&  \+ w\,\left\{\,\sum_k\,(P_k^2 + Q_k^2)\,s_k^2 
\- \left[\,\sum_k\,(P_k^2 + Q_k^2)\,s_k\,\right]^2\,\right\}.\\
&&\label{Ham2}
\ear

Next I wrote Hamilton's equations:

\bar
\no \frac{\partial Q_k}{\partial t} &\=& \frac{\partial}{\partial P_k}\, \cH;\\
 \frac{\partial P_k}{\partial t} &\=& \- \frac{\partial}{\partial Q_k}\, \cH,
\ear
  
\ni and noted that in the linear case ($w = 0$), this system is identical to 
\Sch's equation with the matrix $K$ as the ``kinetic energy" term (in that paper
there was also an external potential term $V(S)$ omitted here).
For the explicit form of these equations, see paper III or Math Appendix 3 below.
I called this a construction of a ``real dynamical system" (RDS); evidentially
it constitutes a 16-dimensional ODE system.

I next introduced the 
linearized or Jacobian dynamical system (JDS) associated to an RDS, defined by:

\bar
\no \frac{\partial \xi_k}{\partial t} &\=& \ph\ph\,\, 
\sjN\,\left\{\, \frac{\partial^2 \cH}{\partial P_k\,\partial Q_j}\,\xi_j \+ 
\frac{\partial^2 \cH}{\partial P_k\, \partial P_j}\,\eta_j \,\right\}\\
 \frac{\partial \eta_k}{\partial t} &\=& \- \sjN\,\left\{\, \frac{\partial^2 \cH}{\partial Q_k\,
\partial Q_j}\,\xi_j \+ \frac{\partial^2 \cH}{\partial Q_k\, \partial P_j}\,\eta_j \,\right\} \label{Jacsystem}
\ear

\ni Here $\xi$ and $\eta$ are real $N$-vectors; in matrix form (\ref{Jacsystem}) can
be written:

\be
\frac{d}{dt}\, \begin{pmatrix}
\xi \\
\eta
\end{pmatrix} \= M(t)\,
\begin{pmatrix}
\xi \\
\eta
\end{pmatrix}.\label{JDSeqns}
\ee

This defines a 16-dimensional, linear ODE system, with time-dependent coefficients.
Often called the ``differential system", the JDS represents 
the motion of a system which approximates the original for a small time interval.

In paper III, the following theorem was proved. 

\begin{quote}
{\bf Theorem} Let $M$ be an even-dimensional ($2N\times 2N$) matrix and

\be
\det\,M < 0.
\ee

\ni Then $M$ has both positive and negative eigenvalues.
\end{quote}

Call it the ``Determinant Criterion for Instability" (DCI).
Since $M$ is a function of the state, the criterion may be satisfied at some
points (states) but not at all.
I will denote the set of wavefunctions (equivalently, the $P_k,Q_k$) for which $M$
satisfies the DCI
as WF(DCI).

\section{Simulations\label{simsection}}

I simulated from the ``spin" or multi-qubit model of paper III with $q = 3$ spins (thus
a 16-dimensional real dynamical system), omitting the external potential.
Figure 1 shows the ``total spin" observable as a function of time for the linear model with $w=0$
(starting with a state with total spin equal to zero); 
clearly, it reflects a
simple oscillation. Fig.2 shows the same observable but with $w = 2.0$.
(The threshold of instability, defined by the condition: $\det\,M < 0$, was $w = 1.15$.)
Now the oscillation is irregular (note the variation of the sequence of local minima of the curve).

I next compared the observables obtained from two sets of initial conditions,
the ``perturbed" being obtained from the ``unperturbed" by making a small rotation
of the variables $P_1$ and $Q_1$. As would be expected, the divergence remained small and
periodic for the linear case, see Fig. 3. By contrast, the divergence is considerably larger,
with nonperiodic growing peaks, see Fig. 4. (The initial logarithm is actually negative infinity,
because the perturbed and unperturbed system both had zero total spin initially. As the
two zero's are computed, then subtracted, the initial difference is due to round-off error.
So ignore
the curves starting points on the left of the figures.) 

Evidently, the perturbed and unperturbed trajectories repeatedly nearly coincided, then diverged,
making it somewhat difficult to see the trend. To smooth out the curves and make the trend obvious,
I computed running maximum deviations (meaning a function that only increases 
when the deviation reaches a new maximum), see Fig. 5. Now a positive slope is apparent up
to (computer) time = 6; since the plot shows the log of the max deviation, 
we have a case of exponential
divergence, not regularily but with quasi-periodic increases.

Also of interest is whether, if the starting wavefunction lies in WF(DCI), does it stay there?
Fig. 6 shows one trajectory; clearly the system departed that set, re-entered, departed, etc.,
in some irregular journey. 

The approach of following two nearby trajectories in order to detect divergence
 suffers from many tribulations and ambiguities.
First, what is meant by ``nearby"? Second, for how long must they be watched? Third,
how much later deviation for how much earlier constitutes ``sensitive dependence on initial
conditions"? 
According to Skokos, the approach was abandoned for another technique: following the evolution of
a tangent vector, which is ``more reliable and efficient" than following two nearby trajectories.
 The idea is to solve the
Hamilton equations together with the deviation equations in (\ref{JDSeqns}), starting with a 
random deviation (tangent) vector, $\{\xi_k,\eta_k\}$. If there is an expanding direction,
with probability one the random vector will have a component in that direction and hence
the norm of the vector will increase. Suppose 

\be
\lim_{t \to \infty}\,\left(\,\frac{1}{t}\,\right)\,\log\left(||\{\,\xi(t),\eta(t)\,\}||\right) 
\= \gamma,
\ee

\ni and $\gamma >0$. Then $\gamma$ is called the ``maximal Lyapunov Characteristic Exponent"
(mLCE) and expresses a rate of exponential divergence of the tangent vector 
which represents a small deviation. 

Skokos notes that there is a serious numerical issue in following the norm of the deviation
over time, if it grows exponentially: overflow. Therefore, the idea is to follow it
for a time period, then renormalize to a unit vector and continue. See \cite{Skokos},
Table 1, p. 34 and preceeding discussion for the detailed algorithm. 

Fig. 7 shows the series estimating the mLCE in a case starting from inside the WF(DIC)
(see the Illustrative Example in Math Appendix 3, after Theorem 3,
for the selection of the intial state), 
and a initial deviation vector chosen as i.i.d. Gaussian and then normalized to unity. 
Fig. 8 shows the same series but in a regime without expected instability (the threshold
with the chosen initial state was $w = 1.15$). Fig. 9 shows a plot of the mLCE
versus parameter `$w$' with the same initial state, averaged over a late time segment. 
 
\begin{figure}
\rotatebox{0}{\resizebox{5in}{5in}{\includegraphics{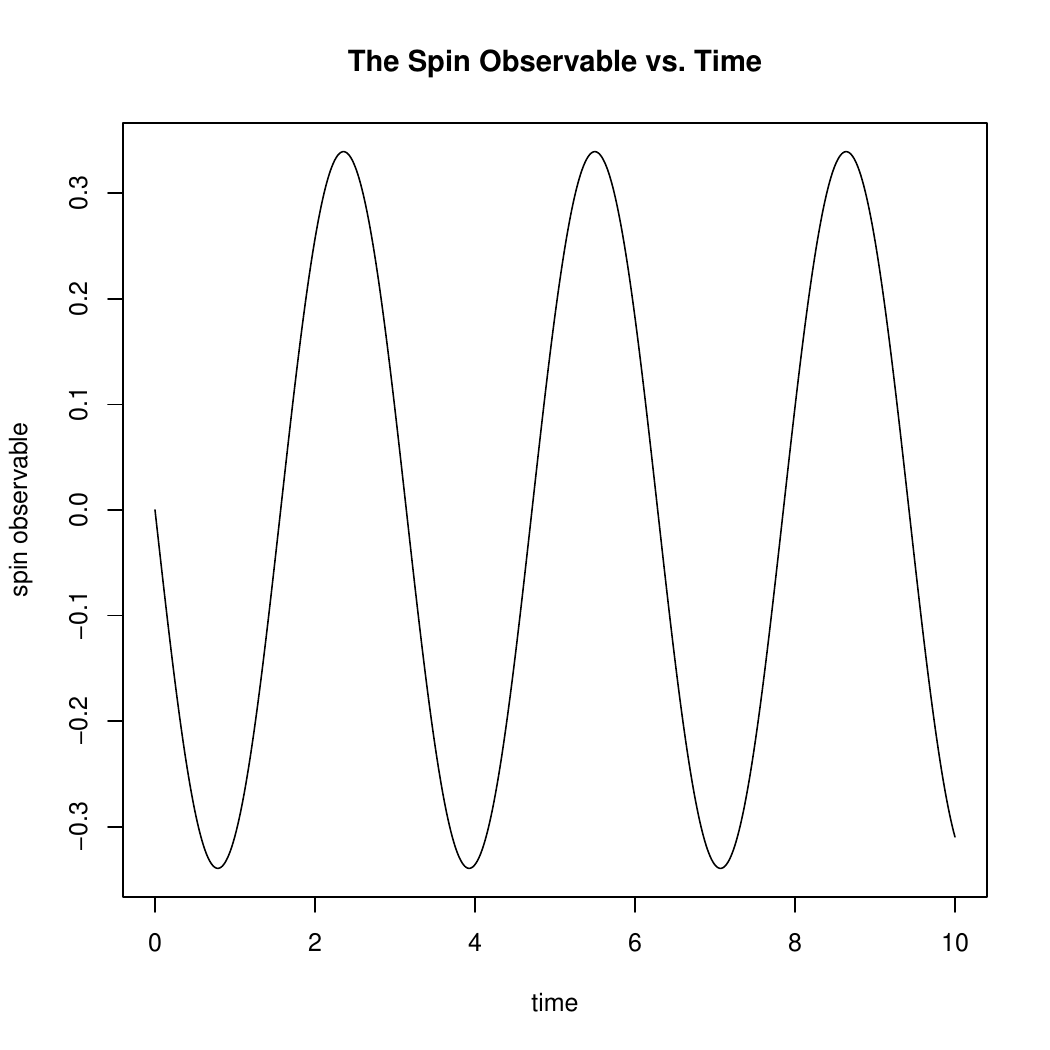}}}
\caption{The Spin Observable vs. time; $w$ = 0.0.}\label{Fig1}
\end{figure}

\begin{figure}
\rotatebox{0}{\resizebox{5in}{5in}{\includegraphics{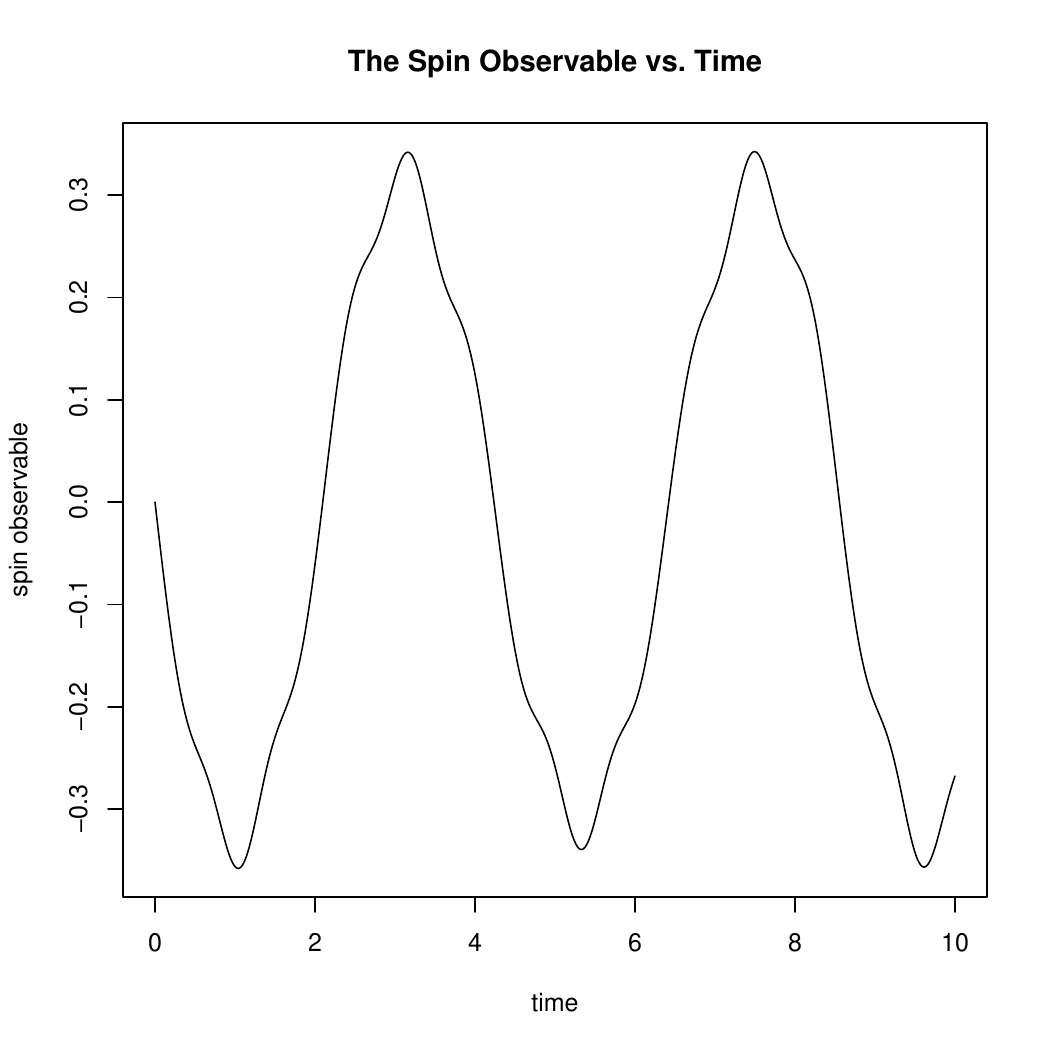}}}
\caption{The Spin Observable vs. time; $w$ = 2.0.}\label{Fig2}
\end{figure}

\begin{figure}
\rotatebox{0}{\resizebox{5in}{5in}{\includegraphics{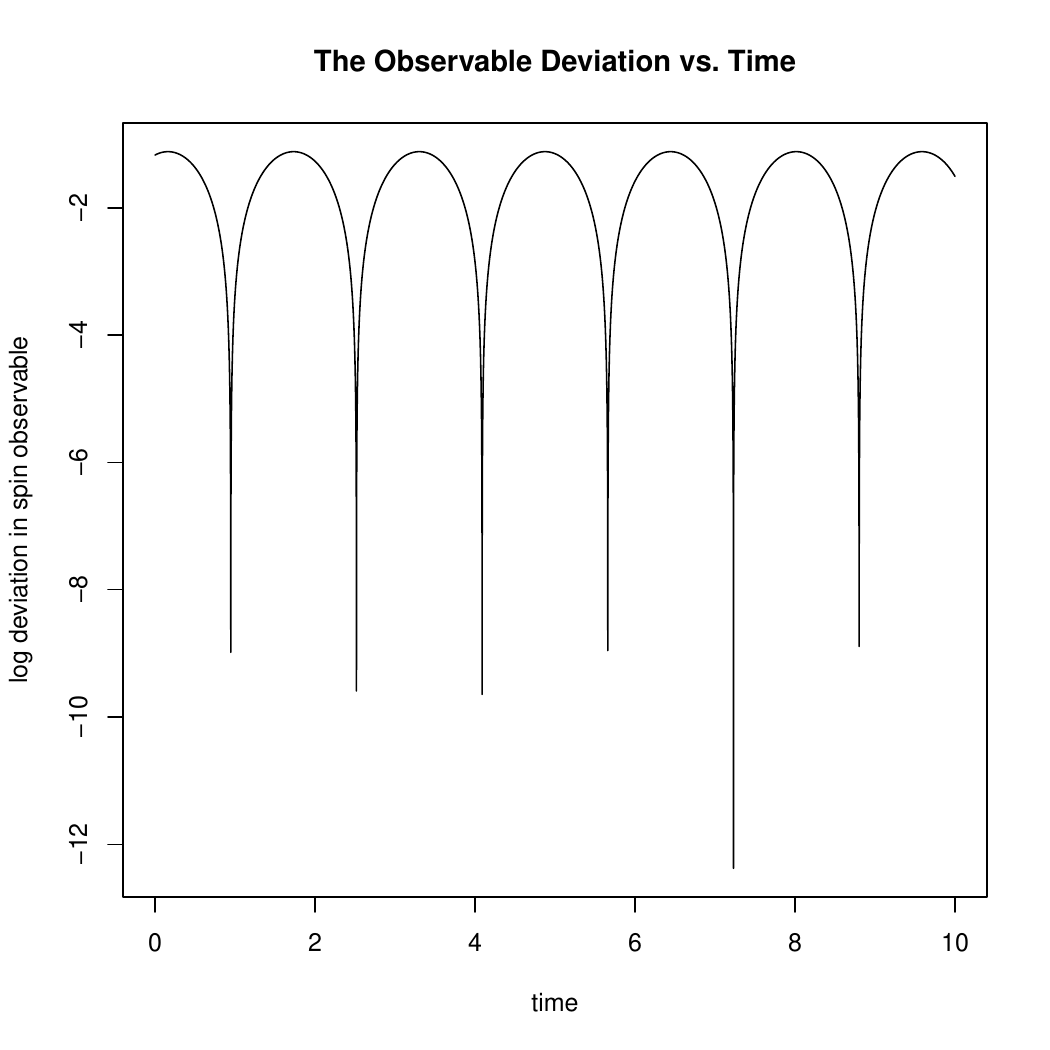}}}
\caption{Log deviation between two trajectories of the spin observable; $w$ = 0.0.}\label{Fig3}
\end{figure}

\begin{figure}
\rotatebox{0}{\resizebox{5in}{5in}{\includegraphics{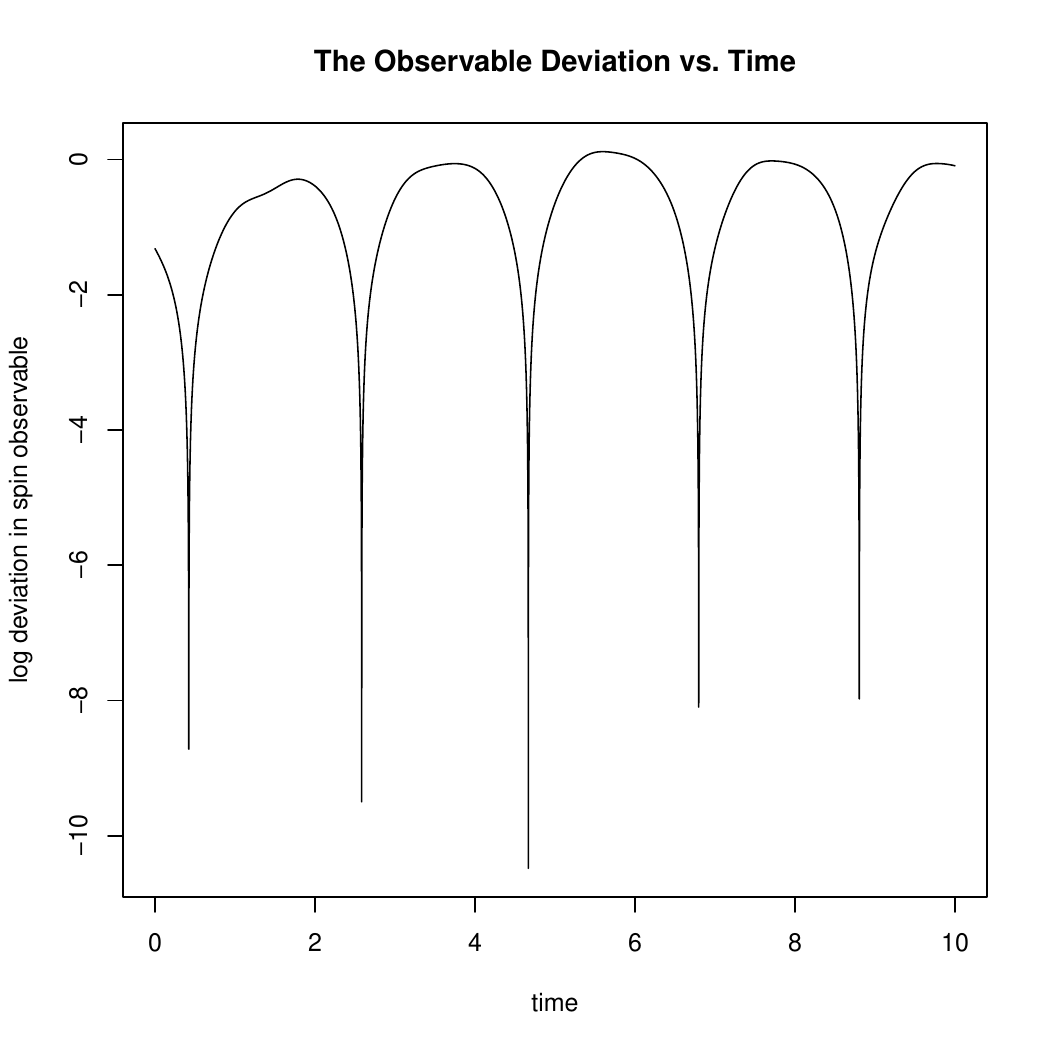}}}
\caption{Log deviation between two trajectories of the spin observable; $w$ = 2.0.}\label{Fig4}
\end{figure}

\begin{figure}
\rotatebox{0}{\resizebox{5in}{5in}{\includegraphics{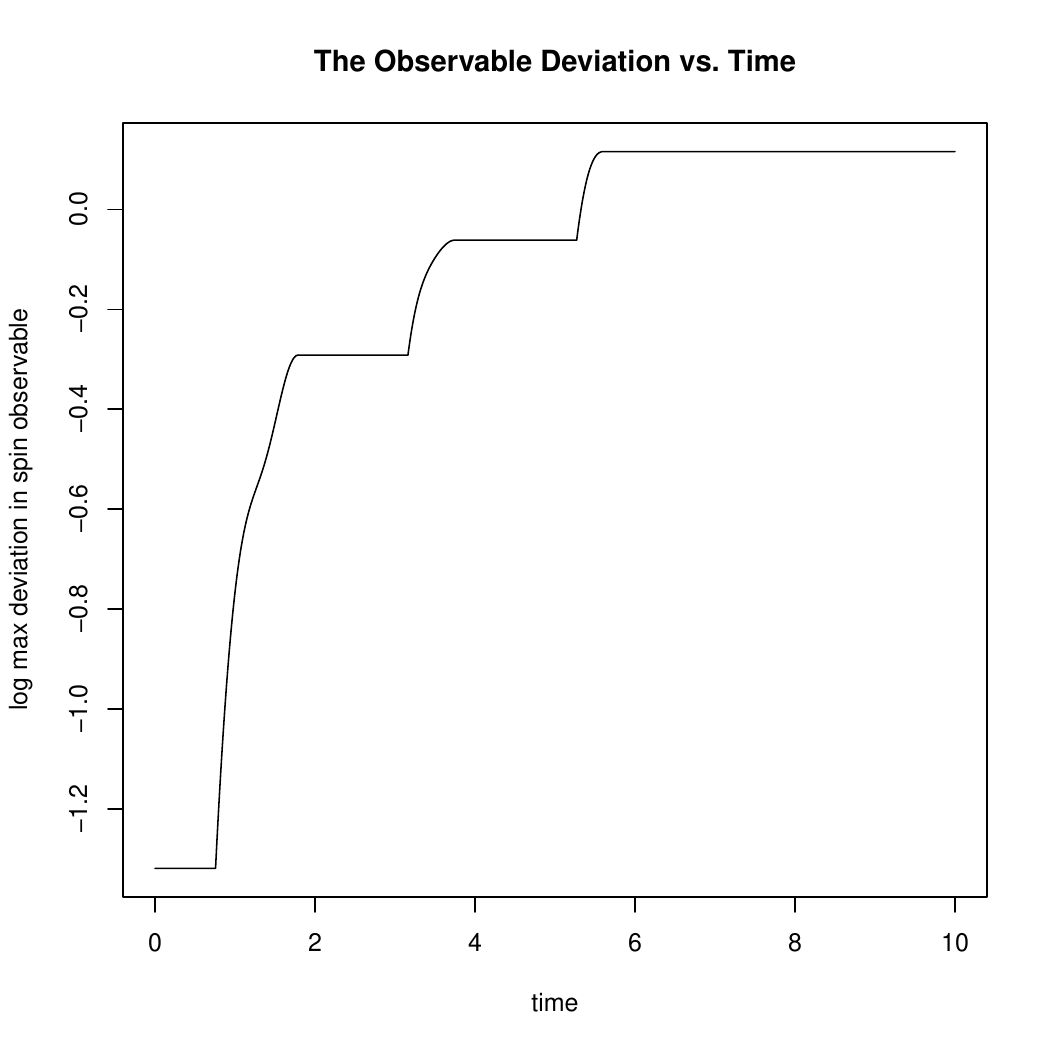}}}
\caption{Log max deviation between two trajectories of the spin observable; $w$ = 2.0.}\label{Fig5}
\end{figure}

\begin{figure}
\rotatebox{0}{\resizebox{5in}{5in}{\includegraphics{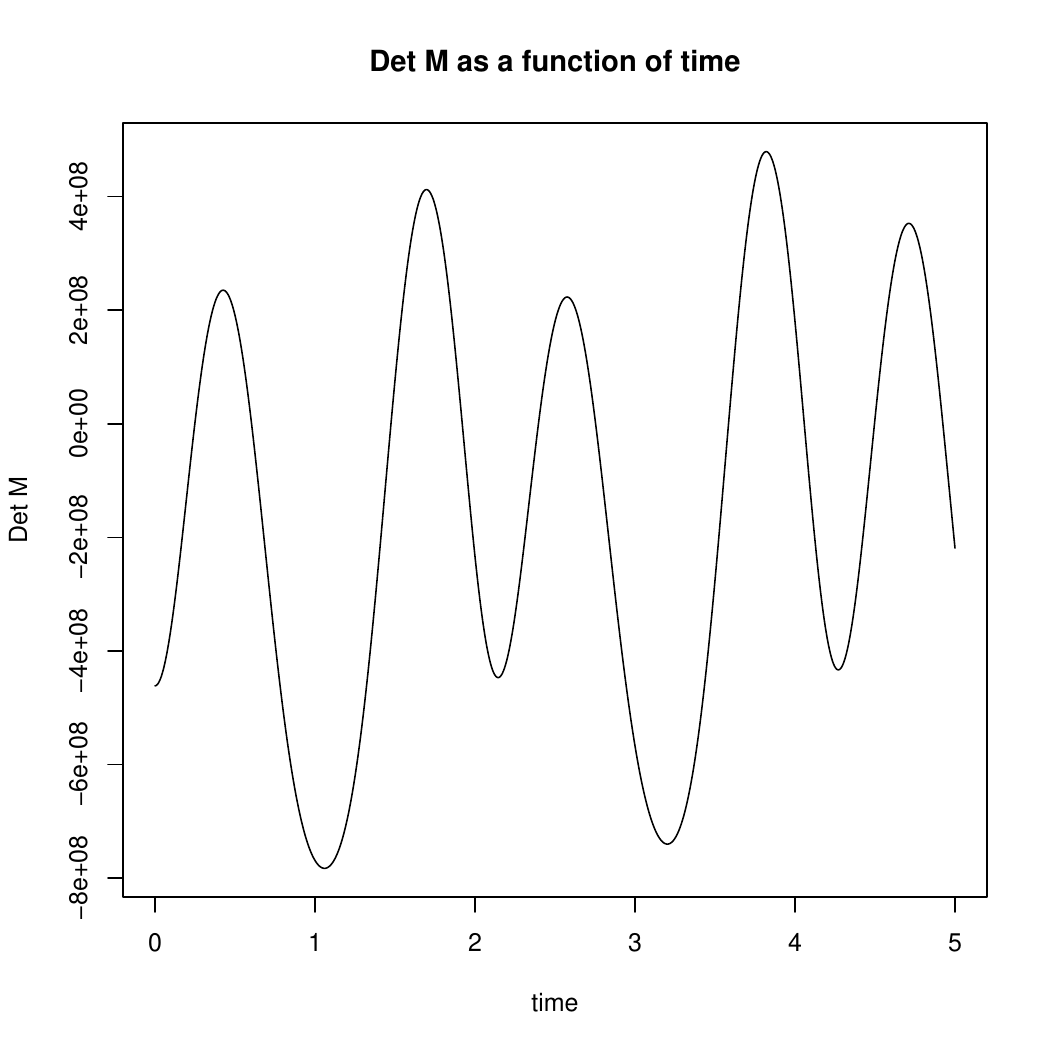}}}
\caption{Det M plotted against computer time; $w$ = 2.0.}\label{Fig6}
\end{figure}

\begin{figure}
\rotatebox{0}{\resizebox{5in}{5in}{\includegraphics{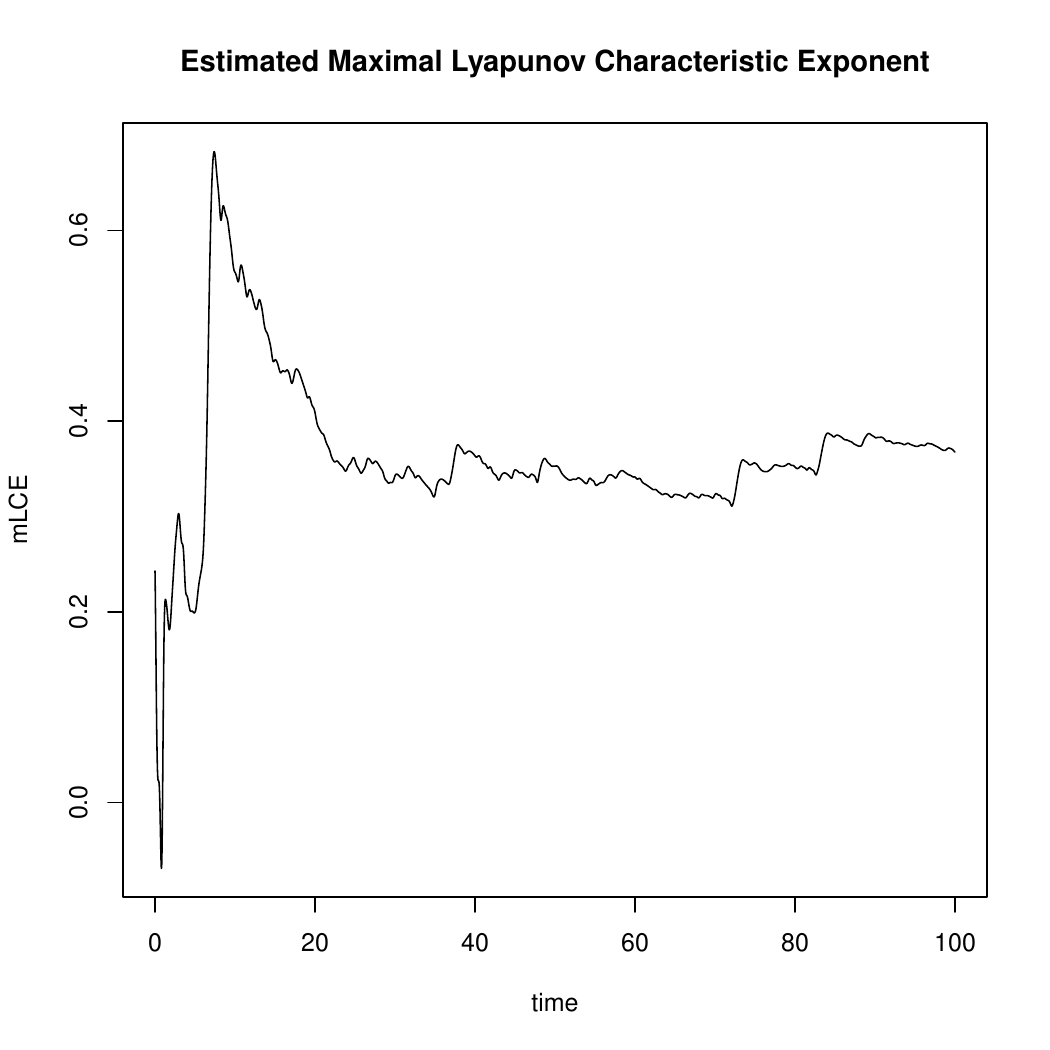}}}
\caption{The series that estimates the mLCE plotted against computer time; $w$ = 3.0.}\label{Fig7}
\end{figure}

\begin{figure}
\rotatebox{0}{\resizebox{5in}{5in}{\includegraphics{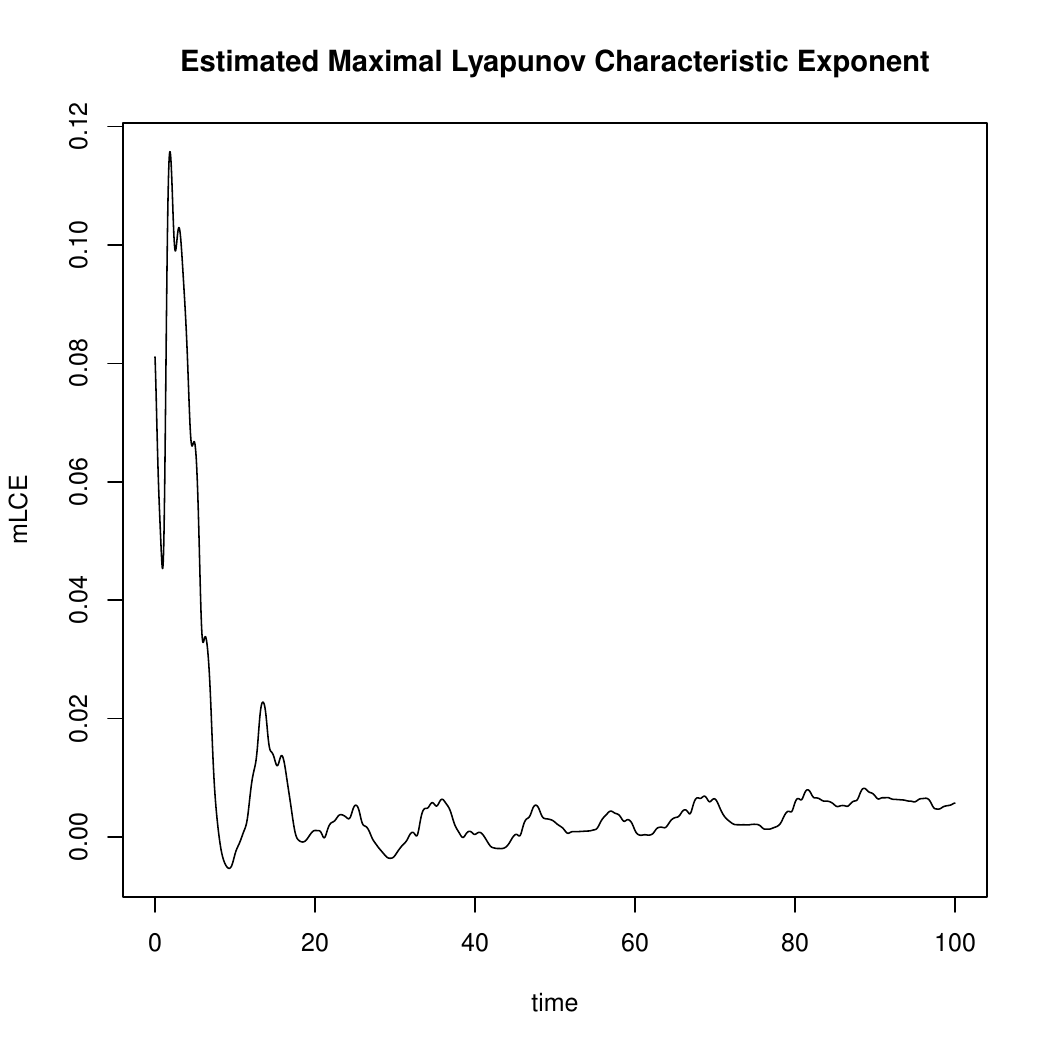}}}
\caption{The series that estimates the mLCE plotted against computer time; $w$ = 1.0.}\label{Fig8}
\end{figure}

\begin{figure}
\rotatebox{0}{\resizebox{5in}{5in}{\includegraphics{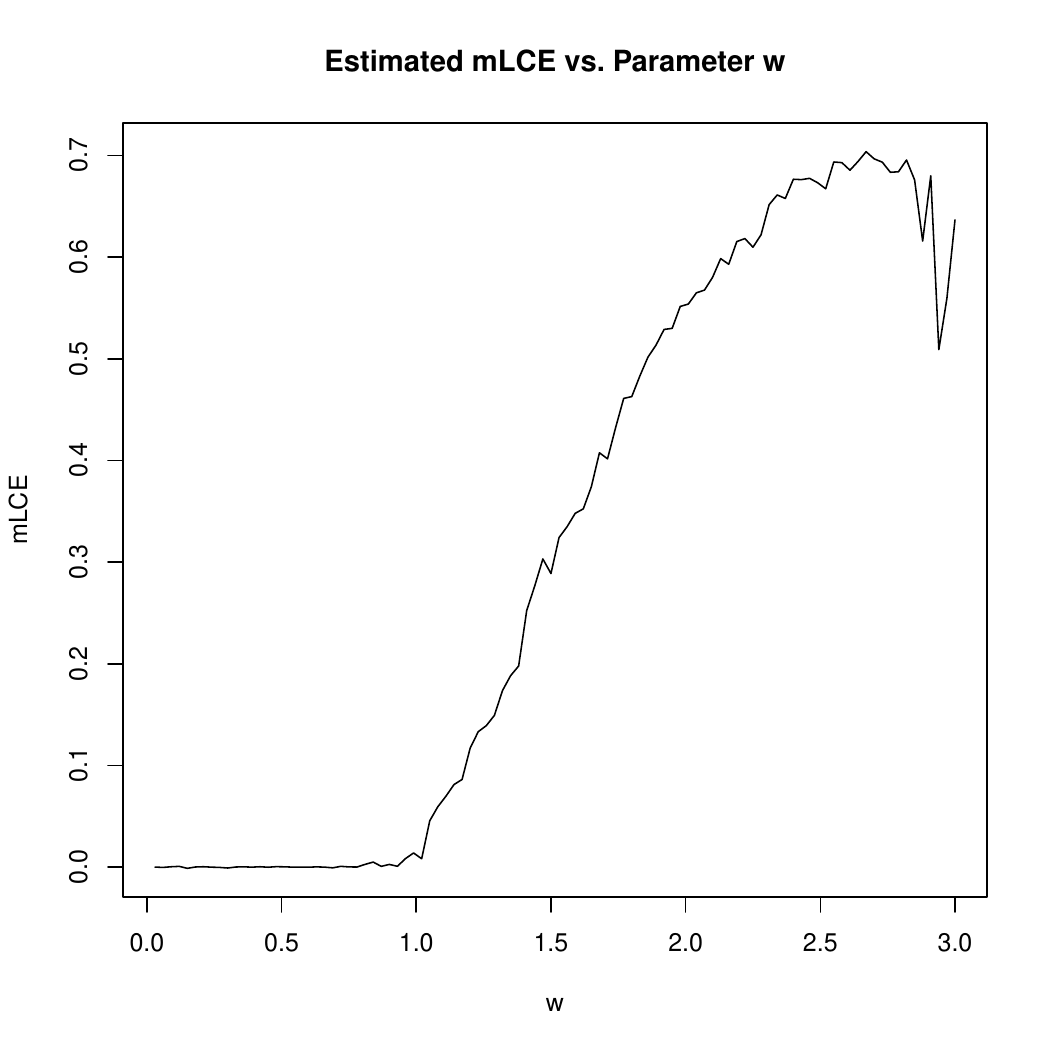}}}
\caption{Estimated mLCE plotted against parameter `w'. The DCI was w = 1.15.}\label{Fig9}
\end{figure}

\pagebreak

\section{Algebraic Results\label{algebrasection}}

The Math Appendicies are devoted to an analysis of the DCI in the context 
of the model from paper III, by exploiting formulas from 
matrix algebra. 
See in particular Theorem 3, its Corollaries, and
the Illustrative Examples near the end, which provide some simple
conditions implying that DCI holds for some wavefunctions. One such set of conditions
can be stated as follows: 

\begin{quote}

(1) The wavefunction has any type of transformation rule under an overall ``spin flip",
i.e., $\sigma_i \to - \sigma_i$, for $1 = 1,2,3$. So $P_k \to \pm P_k$ and $Q_k \to \pm 
Q_k$, with any choice of sign changes allowed;

(2) 

\be \skN \,\left(\,Q_k^2 + P_k^2\,\right) \= 1;
\ee

(3)
 
\be
 \skN\,P_k^2 < 1/4.
\ee
\end{quote}

\ni In other words, the symmetries plus an unbalance between real and imaginary parts of the
 wavefunction suffice for the DCI. 

These conditions are certainly not necessary for the DCI to hold. A broader field of possibilities
is described in the theorems.

\section{Extensions to Continuum Models\label{continuumsection}}

For this work, I chose to treat a small, discrete (3-qubit)
 model out of practical necessity. First, I could represent
the sixteen real numbers describing the state on my computer; second, I had
 a fast (direct), symplectic
(meaning energy-conserving) algorithm (Tao's method) 
to simulate the dynamics; and third, I had a tractable criterion, either expressible
on the yellow pad by algebra, or computed using well-known numerical techniques,
for the existence of an instability.

Finite-qubit or -spin systems are essentially toy models. But if we pass to realistic
continuum models, we lose all these benefits, including the third, 
involving a determinant inequality. I do not know of any obvious or 
plausible extension of that criterion
to the continuum.\footnote{In the area of integral equations, Fredholm did define
a useful determinant, and formulas exist for such in Quantum Field Theory; 
but I could not see how to apply any of them in this context.} 
However, in section \ref{algebrasection} and Math Appendix 2,
 I showed how the determinant condition could be derived from matrix
inequalities. These do have ready analogues in continuum models, in the form of operator 
inequalities.  

I do not want the conclusions of this section to reflect any strange or objectionable features
of such a model.
So I will take as a representative example the simplest known many-body, exactly solvable,
 wavefunction model in the continuum:
`$N$' bodies in one spatial dimension, in a harmonic potential.

The arguments of the wavefunction will be written as:
$(\Ntup)$. 
The analog of the matrix I wrote as `$K$' will now be the operator (setting $\hbar = 1$):

\be
 \ootri \+ V(\Ntup).
\ee

\ni where

\be
\triangle \= \skN\,\frac{\partial^2}{\partial x_k^2}.
\ee

\ni For the potential written as $V(\cdot)$ I take:

\be
V(\Ntup) \= \left(\,\frac{v}{2}\,\right)\,\skN\,x_k^2,
\ee

\ni where `$v$' is a positive parameter.  
  
The eigenvalue-eigenfunction system for this model is well known: in one dimension,
for any `$n$', $n = 0,1,2,...$, let

\be
\psi_n(x) \= \frac{1}{\sqrt{2^n\,n!}}\,\left(\,\frac{m\,\omega}{\pi}\,\right)^{1/4}\,
\exp\left\{\, - m\,\omega\,x^2/2\,\right\}\,H_n\left(\,\sqrt{m\,\omega}\,x\right),\label{hermite}
\ee

\ni where $\omega = \sqrt{v/m}$ and 
$H_n$ denotes the $n$-th Hermite polynomial (of degree $n$ in powers of $x$).
The corresponding eigenvalue is:

\be
\cE_n \= \omega\,\left(\,n + 1/2\,\right).
\ee

For $N$ dimensions, let `$\un$' denote a $N$-tuple
of nonnegative integers, written as if derived from a function $\un:(1,2,...,n) \to 
(R^{+})^N$; i.e., $\un = (n(1),n(2),...,n(N))$.\footnote{To avoid double-subscripts, which LateX
hates.} Then the eigenfunctions are given by:

\be
\psi_{\un}(\Ntup) \= \prod_{k=1}^N\,\psi_{n(k)},
\ee

\ni with corresponding eigenvalue 

\be
\cE_{\un} \= \skN\,\cE_{n(k)}.\label{eigenvl}
\ee

Of the various bundles of conditions for instability derived from algebra, the most promising
for generalizing to continuum models is likely Corollary 2.2 to Theorem 2 of the Math Appendix 2.
First, I define some analogs of quantities appearing there:

\bar
\no f &\=& w\,\ovx\,\left(\,\ovx - 2\,S\,\right);\\
\no \ovx &\=& \skN\,x_k;\\
\no S &\=& \int\,\prod\, d\,x_k\,|\psi(\Ntup)|^2\,\ovx;\\
&&
\ear

The Corollary now translates directly into the criteria:

\begin{quote}

(i) 
\be
\ootri +V(\Ntup) > 0;
\ee
(ii) 
\be
\ootri \+ V(\Ntup) \+ f(\Ntup) > \eta > 0;
\ee
(iii) 
\be 
4\,w\,\int\,\prod\,dx_k\,\left[\,\hbox{Im}\,\psi(\Ntup)\,\right]^2 \,\ovx^2 < \eta;
\ee
(iv) 
\be
f(\Ntup) \leq \left(\,\sigma - 1\,\right)\,\left[\,\ootri + V(\Ntup)\,\right],
\ee
\ni for some $\sigma > 1$;

(v) 
\be
4\,w\,\int\,\pdxk\,\left[\,\Re\,\psi\,\right]\,\ovx\,\left[\,\ootri + V\,\right]^{-1}
\,\ovx\,\left[\,\Re\,\psi\,\right]\, \geq\, \sigma.\label{fifthcondition}
\ee

\end{quote}

Making sense of these criteria requires some explication. First, inequalities involving
operators on the Hilbert space of square-integrable wavefunctions on $R^N$, $\LtRN$, are meant
in the sense:
$A < B$ means that, for all $\psi$ in a suitable dense subset of $\LtRN$,

\be
<\psi|A|\psi>\, =\, \int\,\pdxk\,\psi^*\,A\,\psi\, <\,  \int\,
\pdxk\,\psi^*\,B\,\psi\, =\, <\psi|B|\psi>. 
\ee

I am assuming here that both sides are real numbers, which follows if the operators are symmetric.
A suitable dense subset might be the intersection of sets each in the domain of definitions
of $A$ or $B$. For instance, if $A = \ootri + V$, its domain could be taken to be all
$C^2$ functions decaying appropriately at infinity. Alternatively, the domain could be
all functions in $\LtRN$ whose components along the eigenfunctions: $\{\,<\psi|\psi_{\un}>\,\}$
form a square-summable sequence. (I don't go further into these domain issues, which can assure 
that an operator has a unique extension to be self-adjoint and so define the generator
of a unitary group of transformations, as it is well-covered in the textbooks, e.g., \cite{Bill}.)
For the inverse operator appearing in part (v) above, we can refer to the eigenfunction 
expansion again; i.e., if $\psi = \sum\,<\psi|\psi_{\un}>\,\psi_{\un}$, then

\be
\left[\,\ootri + V\,\right]^{-1}\,\psi \= 
 \sum\,<\psi|\psi_{\un}>\,\cE_{\un}^{-1}\,\psi_{\un},
\ee

\ni which is always finite because $\cE_{\un}$ is bounded below 
(and in fact this inverse operator is bounded).
 
In Math Appendix 4 I prove the

\begin{quote}
{\bf Proposition}

With reasonable conditions on parameters, there exists a state $\psi \in \LtRN$ 
such that the five criteria above hold.
\end{quote} 

I defer asking whether the Proposition implies chaos in a continuum model 
to the next section.

\section{Discussion\label{discsection}}

Chaos in a model with many, even infinitely-many, components
raises the question of what is observable. It might be that e.g., \sdic\ is not reflected
in observable quantities and hence is unable to explain the apparent randomness of
outcomes in some experiment. This qualm is why, in the simulation section,
I looked first at the sum-of-spins (as a surrogate for center-of-mass of some apparatus
part in a realistic, continuum model). On the other hand, all measurement scenarios involve
the creation of an instability in a macroscopic device and its coupling to the microscopic
system observed (because that is how a microscopic signal can be amplified to be detectible).
Thus in such situations we presumably are selecting an observable that is sensitive to
small perturbations, and so we can dismiss the qualm.

Several mathematical questions arise for the discrete models: does the DCI:
$\det M < 0$, actually suffice for chaos, 
perhaps limited to \sdic, and to a finite (perhaps small) time period? 
If so, is the positive eigenvalue of $M$ the same, or related to, the positive Lyapunov exponent?
If the latter is computed using data from that time period only, the DCI would seem
to be sufficient, since over-all expansion should result from infinitesimal
expansion (due to the positive eigenvalue of $M$) ``integrated up".
Is it also necessary?

However, I found, 
when simulating the 3-qubit model, that the system left the WF(DCI) 
and later re-entered, with the sequence repeated. Is this sort of thing generic in this 
class of models? If the Lyapunov exponent is derived from data over longer times, is there
still a way to express it given the history of eigenvalues of $M(t)$?  

Concerning continuum models: I cannot assert that the Proposition of section \ref{continuumsection}
implies the DCI, because I cannot formulate the latter in this context. However, I chose to
investigate finitely-many-qubit models not because of their importance for physics
(although lately they have been popular with those interested in creating a ``quantum computer"),
but as discrete analogs of continuum models. 
One possibility for linking the Proposition to instability and the DCI 
would be to start with a continuum model and then replace $R^N$ with a discrete
lattice (as is often done in calculations in Quantum Field Theory). 
The issue then would
be whether, e.g., Lyapunov exponents remain positive as the lattice spacing is reduced to zero.

There are a handful of solved many-body wavefunction models that incorporate inter-body
forces; see \cite{Hall} for a review. Because of the reduction of the DCI to matrix, or operator,
inequalities made here, I conjecture that chaos could be demonstrated also in these cases.
Most of these models require for their solution identical bodies and exchange symmetry, but 
L. De Carlo and the author partially solved
a model with $N$ light bodies and one heavy body. We were interested in whether
conventional, linear wavefunction physics can explain Brownian Motion (BM). 
The idea would be that BM is purely a consequence of the many frequencies
naturally combining in wavefunction models.
But for the extremely-simplified (one-space-dimensional with repulsive and attractive
forces connecting the light particles and the heavy particle only), the sought-for
behavior did not appear. (Perhaps that was inevitable, as BM is supposed to be related
to viscosity through the diffusion constant, according to Einstein, and viscosity is supposed
to be related to inter-molecular forces in the surrounding fluid.)
But we also noticed that, in the thermodynamic 
ensembles considered, superpositions occurred in which the heavy-body is
dispersed, i.e., is a ``cat". The latter
could be eliminated by incorporating nonlinear terms in the Hamiltonian,
which was the original motivation for studying them (in paper I). If we take that step, 
another possibility would arise: BM 
is a consequence of chaos---and 
the diffusion coefficient is somehow derivable from a positive Lyapunov exponent.
 
Mention of the Lyapunov exponent raises the question of how it depends on model
parameters, especially `$w$' and `$N$'. We have seen that there is a threshold
value of $w$ in a small model; how would it change as $N$ is increased?
(Presumably it will decrease.) In my theory (proposed in paper I) the location
of the ``Infamous Boundary" is fixed by parameters $w$ and $N$ and laboratory energies.
If that location is someday settled by experiments, it will be interesting to learn
whether it is also the threshold to chaos in the observed system.

In 1926, Max Born injected probability into ``quantum physics" apparently out of desperation,
as ``the only possible interpretation" of the wavefunction. He provided no justification for
this radical step away from determinism; and, ignoring mystical or metaphysical conceits,
no one has since. My own views about probability are the following.

Having reviewed the various interpretations of chance or probability, 
I found that I could accept only two: the subjective 
(often called ``Bayesian" after the Rev. Thomas Bayes, an 18th Century English
cleric who dabbled in mathematics), and the dynamical (which can be traced to a book about
probability by Henri Poincar{\'e} published in 1896).\footnote{My views 
on acceptable theories of probability
are explained, with historical references, in a 2013 book, \cite{Wick2013}.} The subjective is the
vulgar mode; it is what we mean when we say ``The chance of the Seahawks winning tonight's game is
50:50." We are not saying anything material about the Seahawks, but only about our opinion.
In other words, probability as a state of mind. Words with such meanings should
never appear on the tongues of physicists during working hours. Physics is supposed to be about
an objective world, not our current opinions. 

The proper interpretation of probability in physics was stated by Poincar{\'e} in his book: 
we resort to
it because the systems we study have instabilities, resulting in \sdic. I therefore
propose that the reason that the upper detector sometimes (in some ``runs" of an experiment)
clicks, while at other times it is the lower one, is because we cannot control all the
details of the wavefunction representing the system (apparatus plus microsystem).
 
You may well ask: didn't Born's hypothesis, however {\em ad hoc}, work rather well?
Perhap so. I discussed whether Born's rule might be derived from models 
exhibiting chaos in paper III, but without any definite conclusion. 
Resolving this issue would require:
introducing more realistic models; determining the Lyapunov exponents for them; 
and learning on what time scales we expect to see apparently random outcomes.

\section*{Math Appendix 1: Matrix identities}

In this section I record many matrix identities and facts I needed for the computations, 
most well-known, some not so much.
The proofs will be displayed for the latter; for the others, 
see any standard linear-algebra text.\footnote{Or Wikipedia. 
Most everything reported here I found
on the Wikipedia page entitled: ``Determinant"; 
on November 10th, 2024. 
Beware of believing everything you find on Wikipedia, 
especially on the history and math pages!
All the other such pages I have inspected contained errors, sometimes egregious. 
So always check before believing anything. The page ``Determinant", at least on that day,
seemed error free.}

In the following, capital Roman letters will denote matrices with real entries.

{\bf Identities and Facts for Square Matrices in General:}

{\em Definition of determinant:}

For any $N\times N$ matrix $X$, $\det(X)$ is expressible as a polynomial, homogeneous and of degree $N$,
in the components of $X$ (see: Leibniz formula for determinants).

{\em Meaning of the determinant:}

$X$ is invertible, meaning a matrix of the same dimension denoted $X^{-1}$ exists and satisfies:

\be
X^{-1} X \= X\,X^{-1} = I_N,
\ee

where $I_N$ denotes the $NN$ identity matrix: $(I_N)_{i,j} = \delta_{i,j}$,
  iff $\det(X) \neq 0$. $X$ is invertible and $\det X \neq 0$ iff 
there exists a null vector of $X$; i.e., an N-vector $\psi$ 
such that $X \psi = 0$.\footnote{It was said in paper III incorrectly
that such a $\psi$ exists provided that $X$ is symmetric, 
but it is a general fact about square matrices.} 

{\em Determinant of a product:}

For any pair of square matricies $X$ and $Y$ of same dimensions, 

\be
\det(XY) \= \det(X)\,\det(Y).
\ee

{\em Determinant of an inverse matrix:}

\be
\det X^{-1} \= \left(\,\det\,X\,\right)^{-1}.
\ee

{\em Determinant of a transposed matrix:}

For any square matrix $X$, $\det(X^t) \= \det(X)$, where $X^t$ denotes the transposed
matrix: $(X^t)_{i,j} = X_{j,i}$.

{\em Determinant of a symmetric matrix:}

A symmetric matrix is one equal to its transpose. Such a matrix has a complete set
of eigenvectors:

\be
X\,\psi_n \= \lambda_n\,\psi_n,
\ee

\ni for $n = 1,2,...,N$. The $\lambda_n$ are the roots (zeroes) of the degree-N polynomial:

\be
\det\left(\,X - \lambda\,I_N\,\right) \= 0.
\ee

The $N$-vectors $\psi_n$ are called ``eigenvectors".
The $\lambda_n$ are called ``eigenvalues" and all are real.
 Any N-vector can be expanded in terms of these eigenvectors. Also:

\be
\det\,X \= \prod_{n=1}^N\,\lambda_n.
\ee

{\em Functions of a symmetric matrix}
Another way to describe the eigenvector-eigenvalue representation of a symmetric matrix is
as follows. There exists an orthogonal matrix $O$, meaning $O^t = O^{-1}$, such that

\be
X \= O^t\,\hbox{diag}(\lambda)\,O,
\ee

\ni where $\hbox{diag}(\lambda)_{i,j} = \delta_{i,j}\,\lambda_j.$

Now given any function of a real variable, say $F(x)$ with domain of definition including all
the $\lambda_n$, one can define

\be
F(X) = O^t\,\hbox{diag}[F(\lambda)]\,O,
\ee

\ni in an obvious notation (meaning replace on the diagonal $\lambda_n$ by $F(\lambda_n)$).
So, for e.g., positive symmetric matrices one can obtain a square-root matrix, and so forth.

The next identity holds for two rectangular matrices: $A$, which is $N\times n$, and $B$, 
which is $n\times N$,
so that both $A\,B$ and $B\,A$ are defined:

{\em Sylvester's determinant theorem (aka the ``Weinstein-Aronszai identity")}:

\be
\det(I_N + A\,B) \= \det(I_n + B\,A).
\ee

{\em A corollary of Sylvester's theorem:}

Given two $N$-vectors $u$ and $v$, define an $N\times N$ matrix called $v\otimes u$ by:
$(v\otimes u)_{i,j} \= v_i\,u_j$. (This matrix is called ``rank one" and has any vector
orthogonal to $u$ as null vector.) Let $X$ be an invertible $N\times N$ matrix. Then:

\be
\det(X + v\otimes u) \= \det(X)\,\left[\,1 + u^t\,X^{-1}\,v\,\right].\label{corrtoSyl}
\ee

\ni Proof: divide out $X$ and apply Sylvesters's theorem. 

{\bf Identities about block matrices:}

In this section let $A$,$B$,$C$ and $D$ be $N\times N$ matrices and define an $2N\times 2N$
matrix by:

\be
M \= \begin{pmatrix}
A & B \\
C & D
\end{pmatrix}
\ee

\ni $M$ is called a ``block matrix".

{\em Multiplying block matrices:}

Let $D$, $E$, $F$, $G$, and $H$ be some more $N\times N$ matrices. Then:

\be
\begin{pmatrix}
A & B\\
C & D
\end{pmatrix}\,\begin{pmatrix}
E & F \\
G & H
\end{pmatrix} \= \begin{pmatrix}
AE + BG & AF + BH \\
CE + DG & CF + DH
\end{pmatrix}
\ee

{\em Transpose of a block matrix:}

\be
\begin{pmatrix}
A & B \\
C & D
\end{pmatrix}^t \=
\begin{pmatrix}
A^t  & C^t \\
B^t & D^t
\end{pmatrix} 
\ee

{\em Reversal of a block matrix (reversing roles of diagonal and off-diagonal blocks):}

Define a $2N\times 2N$ matrix by:

\be 
R \= \begin{pmatrix}
0 & I_N \\
I_N & 0
\end{pmatrix}
\ee

Then:

\be
M\,R \= \begin{pmatrix}
B & A \\
D & C
\end{pmatrix} 
\ee

{\em Determinants of block matrices:}

{\em Determinant of a block matrix with some commutation condition:}
Assume $C\,D = D\,C$, Then

\be
\det\,
\begin{pmatrix}
A & B\\
C & D
\end{pmatrix} = \det\left(\,A\,D - B\,C\,\right).
\ee

{\em Determinant of a reversed matrix:}

\be
\det\left[\,\begin{pmatrix}
B & A \\
D & C
\end{pmatrix}\,\right] \= (-1)^N\,\det\,M.
\ee

\ni Proof: Look at the formula for the reversed matrix.
Then note that, by the formula with the commutation condition:

\bar
\no \det\,R &\=& \det\left( - I_N\,\right)\\
\no &\=& (-1)^N.\\
&&
\ear

{\em Reducing determinants of $2N\times 2N$ block matrices to evaluating determinants of
$N\times N$ matrices:}

Let $A$ be invertible. Then:

\be
\det\,
\begin{pmatrix}
A & B\\
C & D
\end{pmatrix} = \det\,A \,\det\left(\,D - C\,A^{-1}\,B\,\right).
\ee

\ni Proof: 

\be
\begin{pmatrix}
A & B\\
C & D
\end{pmatrix}\, 
\begin{pmatrix}
A^{-1} & - A^{-1}\,B\\
0 & I_N
\end{pmatrix}\=
\begin{pmatrix}
I_N & 0\\
C\,A^{-1} & - C\,A^{-1}\,B + D
\end{pmatrix}
\ee

In the case appearing in this paper, in matrix M the 
matrices $A$ and $D$ are rank-one and so not invertible, while matrices $B$ and $C$ may be.
So we make a reversal:

\be
\det\,
\begin{pmatrix}
B & A \\
D & C
\end{pmatrix} = \det\,B \,\det\left(\,C - D\,B^{-1}\,A\,\right).
\ee

\ni Thus obtaining the useful formula: Assuming 
$B$ is invertible,

\be
\det\,M \= (-1)^N\,\det\,B \,\det\left(\,C - D\,B^{-1}\,A\,\right).
\ee

\section*{Math Appendix 2: Conditions implying instability.}

\def\diagf{\hbox{diag}(f)}

We have to study conditions for which the determinant of the $2N\times 2N$ matrix:

\be
M \= \begin{pmatrix}
A & B \\
C & D
\end{pmatrix}
\ee

\ni is negative. We can write the four matrices simply as:

\bar
\no A &\=& - u\otimes\,v;\\
\no D &\=& - A;\\
\no B &\=& E - u\otimes\,u;\\
\no C &\=& - E + v\otimes\,v;\\
&&
\ear

\ni Here $u$ and $v$ are N-vectors and 

\be
E \= K + \diagf,
\ee

\ni where $K$ denotes the matrix derived from the linear terms in the Hamiltonian, 
assumed to be symmetric;
$f$ is another N-vector; and $\diagf$ is an $N\times N$ diagonal matrix with
entries: $\diagf_{i,i} = f_i$ for $i=1,2.,,,N$. Note that $B$ and $C$ are symmetric,
but not $A$ or $D$.

Our first assumption is that $E$ is invertible. Indeed, the obvious scenario is that $E > 0$. 
(For symmetric matrices, inequalities such as $X > \eta$, for some number $\eta$,
 can be interpreted as: $\psi^t\,X\,\psi >
\eta\,\psi^t\,\psi$ for any N-vector $\psi$, or equivalently that every
eigenvalue of $X$ is greater than $\eta$.)
We can choose to study models with $K > 0$; but $f$ may not have all positive entries. 
(E.g., in the model of paper III.)
As $f$ has a prefactor of `$w$', 
the coupling constant in the nonlinear theory, this scenario may require an upper bound on $w$.

We will rely on exploiting the formula derived in Math Appendix 1:

\be
\det\,M \= (-1)^N\,\det\,B \,\det\left(\,C - D\,B^{-1}\,A\,\right).\label{formula}
\ee

\ni which was proven assuming matrix $B$ is invertible. The following theorem applies in our
context:

\begin{quote}
{\bf Theorem 1.} Let 

\be
B \= E - u\otimes u,
\ee

\ni and assume $E$ is symmetric and invertible and

\be
u^t\,E^{-1}\,u \neq 1.\label{Euineq}
\ee

\ni Then $B$ is invertible and

\be
B^{-1} \= E^{-1} + \left(\,1 - u^t\,E^{-1}\,u\,\right)^{-1}\,
E^{-1}u\otimes\,E^{-1}u.\label{Binvformula}
\ee
\end{quote}

Proof: Define for convenience:

\be
\rho \= \left(\,1 - u^t\,E^{-1}\,u\,\right)^{-1}.
\ee

Multiplying the claimed expression for $B^{-1}$ by $B$ gives:

\bar
\no && \left(\,E - u\otimes u\,\right)\,E^{-1} + 
\rho\,\left(\,E - u\otimes u\,\right)\,E^{-1}\,u\otimes E^{-1}\,u \= \\  
 \no && I_N - u\otimes\, u^t\,E^{-1} + \rho\,
\left(\,1 - u^t\,E^{-1}\,u\,\right)\,u\otimes\,E^{-1}\,u \= I_N.\\
&&
\ear

QED.

A simple scenario implying condition (\ref{Euineq}) is given by:

\begin{quote}
{\bf Lemma 1} Assume $E$ is symmetric, 
$E \geq \eta$ for some number $\eta > 0$, and $u^t\,u < \eta$.
Then 

\be
u^t\,E^{-1}\,u < 1.
\ee
\end{quote}

\ni Proof: plug $\psi = E^{-1/2}\,u$ into the inequality:

\be
\psi^t\,E\,\psi \geq \eta\,\psi^t\,\psi.
\ee

\ni QED.

I note that the square-root of a positive matrix always exists, and can be constructed by
diagonalizing as mentioned in Math Appendix 1.

We now can begin on evaluating formula (\ref{formula}) and search for sufficient conditions
implying $\det M < 0$. By Sylvester's theorem we have:

\be
\det\,B \= \det E\,\left(\,1 - u^t\,E^{-1}\,u\,\right).
\ee

For the second factor we can compute it using:

\be
C - D\,B^{-1}\,A \= - E + \left[\,v + \left(\,v^t\,B^{-1}\,u\,\right)\,u\,\right] \,\otimes\,v.
\ee

\ni so again by Sylvester's theorem:

\be
\det\,\left(\,C - D\,B^{-1}\,A\,\right) \= \det\left(\, - E\,\right)\,\left(\,
1 + v^t\,\left(\, - E\,\right)^{-1}\,  \left[\,v + \left(\,v^t\,B^{-1}\,u\,\right)\,u\,\right]
\right) 
\ee

Combining these formulas leads to:

\be
\det\,M \= \left(\,\det E\,\right)^2\,
\left(\,1 - u^t\,E^{-1}\,u\,\right)\,
\left[\, 1 - v^t\,E^{-1}\,v - v^t\,B^{-1}\,u\,v^t\,E^{-1}\,u\,\right].
\ee

Thus we can state the

\begin{quote}
{\bf Theorem 2.} (Sufficient conditions for the instability)
Let $E$ be symmetric and invertible and either of the pair of inequalities:

\bar
\no u^t\,E^{-1}\,u &<& 1 \ph \hbox{(top 1)}\\
\no &>& 1;\ph \hbox{(bottom 1)}\\
\no v^t\,E^{-1}\,v + \left(\,v^t\,B^{-1}\,u\,\right)\,v^t\,E^{-1}\,u &>& 1\ph \hbox{(top 2)}\\
\no &<& 1\ph \hbox{(bottom 2)};\\
&&
\ear

\ni holds. Then:

\be
\det M < 0.
\ee

\end{quote}

In the following I will concentrate on elucidating conditions implying inequalities 
top 1 and top 2 in Theorem 2 hold.

Remark: suppose $u = v =0$, which holds in the model of paper III if the coupling constant
with the nonlinear term, written `$w$', vanishes. Then

\be
\det M = \left(\,\det E\,\right)^2 > 0.
\ee

\ni This result is consistent with the observation made in paper III that conventional,
linear wavefunction theory has no instabilities.

We next exploit Theorem 2 and (\ref{Binvformula}). Let 

\bar
\no x &\=& v^t\,E^{-1}\,v;\\
\no y &\=& v^t\,E^{-1}\,u;\\
\no z &\=& u^t\,E^{-1}\,u.\\
&&
\ear

Assuming $z < 1$ we compute using (\ref{Binvformula}): 

\be
v^t\,B^{-1}\,u \= \frac{y}{1 - z}.
\ee

\ni We can then write:

\be
\hbox{top 2:}\ph \Leftrightarrow x + \frac{y^2}{1 - z} > 1.
\ee

We can now derive some simple necessary or sufficient conditions for the desired result.
Simplest is the sufficient conditions: $z < 1$ and $x > 1$ or:

\begin{quote}
{\bf Corollary 2.1}.Sufficient conditions for 
$\det M < 0$ and instability are: $E$ is symmetric, invertible, and

\bar
\no u^t\,E^{-1}\,u &<& 1;\\ 
\no v^t\,E^{-1}\,v &>& 1.\\ 
&&
\ear
\end{quote}

\begin{quote}
{\bf Corollary 2.2.} The following conditions are also sufficient to conclude instability:

(i) $K$ is symmetric and $K > 0$;

(ii) $K + \diagf \geq \eta$ for some $\eta > 0$;

(iii) $u^t\,u < \eta$;

(iv) $\diagf \leq \left(\,\sigma - 1 \,\right)\,K$, for some $\sigma > 1$;

(v) $v^t\,K^{-1}\,v > \sigma$.

\end{quote}

\ni Proof: The following lemma complements Lemma 1:

\begin{quote}
{\bf Lemma 2}. Let $X$ be a symmetric matrix enjoying: $ 0 < X < \delta$, for some $\delta > 0$.
Let $\xi$ be an N-vector. Then $\xi^t\,\xi > \delta$ implies $\xi^t\,X^{-1}\,\xi > 1$.
\end{quote}

Proof: We have assumed

\be
0 < \theta^t\,X\,\theta < \delta\,\theta^T\,\theta
\ee

\ni for all N-vectors $\theta$. Substitute: $\theta = X^{-1/2}\,\xi$ above. QED

Proof of Corollary 2.2: We can write:

\be
E \= K^{1/2}\,\left(\,I_N + K^{-1/2}\,\diagf\,K^{-1/2}\,\right)\,K^{1/2},
\ee

\ni from which there follows

\be
E^{-1} \= K^{-1/2}\,\left(\,I_N + K^{-1/2}\,\diagf\,K^{-1/2}\,\right)^{-1}\,K^{-1/2}.
\ee

\ni and so we have:

\be
v^t\,E^{-1}\,v \= \left(\,K^{-1/2}\,v\,\right)^t\,
\left(\,I_N + K^{-1/2}\,\diagf\,K^{-1/2}\,\right)^{-1}\,K^{-1/2}\,v.
\ee

\ni We apply Lemma 2 with 

\be
X \= I_N + K^{-1/2}\,\diagf\,K^{-1/2}.
\ee

\ni and $\delta = \sigma$. The hypothesis is:

\be
\xi^t\,\xi + \xi^t\,K^{-1/2}\,\diagf K^{-1/2}\,\xi \leq \sigma\,\xi^t\,\xi,
\ee

\ni which, substituting $\xi = K^{1/2}\,\theta$, can be rewritten as hypothesis (iv)
of the theorem. QED.

Some simple necessary conditions for top 2 
arise assuming $E > 0$ and applying Cauchy-Schwarz to $y$:

\bar
\no |y| = |v^t\,E^{-1}\,u| &=& |\left(\,E^{-1/2}\,v\,\right)^t\,\left(\,E^{-1/2}\,u\,\right)|\\
\no &\leq& \sqrt{v^t\,E^{-1}\,v}\,
\sqrt{u^t\,E^{-1}\,u};\\
&&
\ear

\ni hence:

\be
|y| \leq \sqrt{x\,z}.
\ee

\ni Therefore

\be
x + \frac{y^2}{1 - z} \leq \frac{x}{1 - z}.
\ee

\ni and so we must have $x + z > 1$ or:

\begin{quote}
{\bf Corollary 2.3}. A necessary condition for employing top 2, assuming $E > 0$ and $z < 1$ is:

\be
v^t\,E^{-1}\,v + 
 u^t\,E^{-1}\,u > 1 
\ee

\end{quote}

\section*{Math Appendix 3: Fulfilling the instability conditions in specific instances}
 
Stating sufficient conditions that are not also necessary raises the question of consistency:
are there any instances where they are satisfied? I next address this question
by studying specific cases.
I specialize the model to that of paper III, which imposes an additional constraint
on the vectors $u$ and $v$ and makes explicit the vector $f$. These are given in terms of
that model by:

\bar
\no f_j &\=& w\,s_j\,\left\{\,s_j - 2\,\sum_{k=1}^N\,s_k\,\left(\,Q_k^2 + P_k^2\,\right)
\,\right\};\\
\no v_k &\=& 2\,\sqrt{w}\,s_k\,Q_k;\\ 
\no u_k &\=& 2\,\sqrt{w}\,s_k\,P_k.\\ 
&&
\ear

Here $Q_k$ and $P_k$ are the real and imaginary components of the wavefunction and so
satisfy

\be
\no \sum_{k=1}^N\,\left(\,Q_k^2 + P_k^2\,\right) \= 1.
\ee

So we switch from $u,v$ to $Q,P$ in terms of which we find for the relevant quantities:

\bar
\no x &\=& 4\,w\,Q^t\,\diags\,E^{-1}\,\diags\,Q;\\
\no y &\=& 4\,w\,Q^t\,\diags\,E^{-1}\,\diags\,P;\\
\no z &\=& 4\,w\,P^t\,\diags\,E^{-1}\,\diags\,P.\\
&&
\ear.

As in Math Appendix 2, sufficient conditions are: 

\bar
\no && (a) E > 0;\\
\no && (b) \ph z < 1;\\
\no && (c) \ph (x-1)\,(1 - z) + y^2 > 1;\\
&&\label{suff}
\ear 

\ni or, an alternative to (c):

\be
(c') \ph x > 1.
\ee

In order to find instances where some sufficient conditions are fulfilled,
it is useful to write:

\bar
\no E &\=& K + w\,J;\\
\no J &\=& \diagss - 2\,S\,\diags;\\
\no S &\=& \sum_{k=1}^N\,s_k\,\left(\,Q_k^2 + P_k^2\,\right).\\
&&
\ear

We now pursue the following strategy: to satisfy (a) assuming that $K \geq 0$, 
it suffices to know that $J > 0$. To obtain conditions under which (b) and (c) or (c')
hold, we will investigate the limit as $w \to +\infty$.

For the rest of this section I will refer to the spin model of paper III to derive various
facts needed for the proof of the instability conditions, although it will be apparent
that these facts will hold in many other models.

Consider a collection of `$q$' classical spins taking values
$\pm 1/2$, where $q$ is a positive, odd integer. Then there are $N \= 2^q$ possible configurations.
In this instance, a subscript index such as `$k$' or `$n$' stands for one of the configurations,
and let $s_k$ be the total (sum) of spins of configuration `$k$'.\footnote{In paper III,
I unfortunately used $s_k$ with different meanings in sections 3 and 4. 
In section 4 it meant
an individual spin, but in section 3 a sum of spins of a configuration. 
Equation (35) was meant to clarify the situation, 
but different notation should have been employed in the different sections.} 
Note that $s_k \neq 0$
for any `$k$' (because `$q$' is odd). Also, as in that model, I assume $K > 0$.

To satisfy (a) of (\ref{suff}) it suffices to know that $J > 0$, equivalent to asking:

\be
2\,s_n^{-1} S < 1,\ph \hbox{ for all } n = 1,2,..,N.
\ee  

Since this inequality is vacuous if $s_n < 0$ for some n, we need only require 
that it hold for the minimum total spin, which with $q$ spins-1/2 is + 1/2,
so we require $ S < 1/4$.

{\bf Illustrative Example:} I switch for this example to spin notation.
So let $\sigma = \{\,\sigma_r = \pm 1/2\,:\,r = 1,2,...q\,\}$, be a spin configuration. Let

\be
\overline{\sigma} \= \sum_{r=1}^{q}\,\sigma_r
\ee

\ni be its total spin. Let `${\cal E}$' denote expectation relative to the flat measure
on spin configurations; i.e., for any function $F(\sigma)$:

\be
\cE\,F \= \left(\,\frac{1}{2^q}\,\right)\,\sum_{\sigma}\,F(\sigma). 
\ee

\ni (Alternatively, $\cE$ can be defined as the product of 1/2-1/2 weights on the spins.)
Now let there be two functions, $f(\sigma)$ and $g(\sigma)$ which have either symmetry under
flipping all the spins:

\bar
\no f(\tau\,\sigma) &\=& \pm\,f(\sigma);\\
\no g(\tau\,\sigma) &\=& \pm\,g(\sigma);\\
&&
\ear

\ni where $(\tau\,\sigma)_r = -\sigma_r$ and any choices of the signs are acceptable. Then let

\be
Q(\sigma) \= \frac{f(\sigma)}{ \sqrt{2\cE\,f^2}}; \ph 
P(\sigma) \= \frac{g(\sigma)}{\sqrt{2 \cE\,f^2} }. 
\ee

The result is:

\bar
\no \cE\,\left\{\,Q(\sigma)^2 + P(\sigma)^2\,\right\} &\=& 1;\\ 
\no \cE\,\left\{\,Q(\sigma)^2 + P(\sigma)^2\,\right\}\,\overline{\sigma} &\=& 0.\\ 
&&
\ear

Translating back to previous notation, where $k \leftrightarrow \sigma$, we obtain
both the constraint and $S = 0$  
implying $J = \diagss \geq 0$.

I now return to parts (b) and (c) of conditions in (\ref{suff}). For `$x$' we have:

\be
x \= 4\,Q^t\,\diags\,\frac{w}{K + w\,J}\,\diags\,Q.
\ee

Letting $w \to \infty$ we obtain, assuming $S < 1/4$:

\bar
\no x(\infty) &\=& 4\,Q^t\,\diags\,J^{-1}\,\diags\,Q.\\
\no &\=& 4\,\skN\,Q_k^2\,\left(\,\frac{s_k}{s_k - 2\,S}\,\right),\\
&&
\ear

\ni and similarily, in the limit:

\bar
\no y(\infty) &\=& 4\,\skN\,Q_k\,P_k\,\left(\,\frac{s_k}{s_k - 2\,S}\,\right),\\
\no z(\infty) &\=& 4\,\skN\,P_k^2\,\left(\,\frac{s_k}{s_k - 2\,S}\,\right).\\
&&
\ear

In terms of these quantities we can state the following:

\begin{quote}
{\bf Theorem 3}. Assume, in the spin model:
 
\bar
\no && \skN\,\left(\,Q_k^2 + P_k^2\,\right) \= 1;\\
\no && S \= \skN\,s_k\,\left(\,Q_k^2 + P_k^2\,\right) < 1/4;\\
\no && \ph z(\infty) < 1;\\
\no && \ph [x(\infty)-1]\,[1 - z(\infty)] + y^2(\infty) > 1;\ph \hbox{or:}\\
\no && \ph x(\infty) > 1.\\
&&
\ear

Then there exists a $w^* < + \infty$ such that, for $w > w^*$, 

\be
\det\,M < 0.
\ee
\end{quote}

{\bf Illustrative Example, continued}: Sufficient conditions are
the spin-flip symmetries assumed there and:

\bar
\no && \skN \,\left(\,Q_k^2 + P_k^2\,\right) \= 1;\\
\no && \skN\,P_k^2 < 1/4.\\
&&
\ear

\ni In other words, the symmetries and an unbalance between real and imaginary parts of the
wavefunction. Considering the spin-flip symmetries: there could be no sign change,
or both real and imaginary parts change sign, 
or one component changes sign and the other doesn't.

{\bf Remark}: There is an obvious symmetry of the model 
in which $Q_k$ and $P_k$ can be interchanged.

{\bf Remark}: The theorem involves conditions under which some function 
of $w$, call it $F(w)$, has a limit at infinity which is, e.g., $F(+ \infty) > 1$. 
If the function is continuous, as is true here,
there must be a finite value
of `$w$', call it `$w^*$', for which $w > w^*$ implies $F(w) > 1$. It is given by:

\be
w^* \= \sup\,\left\{\,w: F(w) = 1\,\right\}.
\ee

For several functions considered jointly, we take the largest such value of $w^*$.

Formally, the derivative of $F(w)$ with respect to $w$ is given by, e.g., for the case of $x$:

\be
F'(w) \= 4\,Q^t\,\diags\,\left[\,\frac{K}{(K + w\,J)^2}\,\right]\,\diags\,Q,
\ee

\ni and is positive if $K$ is, so that we could obtain $w^*$ as the unique solution of $F(w) = 1$.
If $K$ and $J$ commute, which is unlikely, this argument is valid, but not otherwise.

\section*{Math Appendix 4: A continuum wavefunction fulfilling the conditions in the Proposition}

I'll first show the parameter and wavefunction
restrictions so as to be able to demonstrate that they can be satisfied; the proofs of
the inequalities follow.

(i) Is fulfilled, for any $\psi$; in fact,

\be
\ootri \+ V \geq \omega\,N/2 \= \cEz.
\ee

(ii) It suffices to require: 

\be
w \leq \frac{\epsilon\,\cEz}{S^2} < \frac{\cEz}{S^2} = 
\frac{\omega\,N}{2\,S^2}.\label{reqone} 
\ee

(iii) We can assume that the wavefunction is real.

(iv) It suffices if:

\be
w \< \hbox{min}\,\left\{\,\frac{(\sigma - 1)\,v}{2\,N},\ph 
\left\{\,\frac{2\,(\sigma-1)\,\delta^3\,}{2\,m}\,\right\}^{1/4}
\,\left[\,\frac{1}{|S|}\,\right]\,\right\}.\label{reqtwo}
\ee

\ni where $\delta = (\sigma-1)\,v/2 - w\,N$.

(v) 

Define an operator on $\LtRN$ by:
 
\be
\Omega \= \ovx\,\left[\,\ootri + V\,\right]^{-1}
\,\ovx.
\ee

Then it suffices if:

\be
w \Geq \frac{\sigma}{4\,<\psi\,|\,\Omega\,|\,\psi>}.\label{reqthree}
\ee

Below I will show that there are wavefunctions with

\be
<\psi\,|\,\Omega\,|\,\psi> \Geq C\,N\,(N+1).
\ee

I will return to the proof of the Proposition of section \ref{continuumsection}
after first proving these claims.

Proof of (\ref{reqone}): Given the lower bound above, it suffices to show that

\be
f \geq - \epsilon\,\cEz
\ee

\ni with $\epsilon < 1$.
Clearly, $f \to \infty$ as $\ovx^2 \to \infty$; so, looking for the minimum we find
$\ovx = S$ and

\be
f \geq - w\,S^2.
\ee

Proof of (\ref{reqtwo}): Let

\be
G(\Ntup) \= w\,\ovx\,\left\{\,\ovx - 2\,S\,\right\} - (\sigma-1)\,v/2\,\skN\,x_k^2.
\ee

\ni We have to find conditions under which:

\be
G(\Ntup) \leq \,\left(\,\frac{\sigma - 1}{2\,m}\,\right)\,\left(\, - \triangle\,\right)
\ee

\ni as operators on $\LtRN$. To begin, complete the square in the quadratic form $G$,
which yields:

\bar
\no G &\=& \sjkN\,A_{j,k}\,(x_j - b_j)\,(x_k - b_k) + B;\\
\no A\,b &\=& w\,S\,1_N;\\
\no B &\=& - b^t\,A\,b.\\
&&\label{ABeqns}
\ear

\ni The matrix $A$ is given by:

\be
A_{j,k} =
\begin{cases}
 -(\sigma - 1) + w,& \hbox{if}\ph j = k;\\
w,& \hbox{otherwise}.
\end{cases}
\ee

The eigenvalues of the symmetric matrix $A$ are the same as those of the matrix
$w\,1_N\otimes 1_N$ translated by $-(\sigma - 1)v/2$:

\be
\det\left(\,A - \lambda\,I\,\right) \= \det\left(
\,w\,1_N\otimes 1_N - \tilde{\lambda}\,I_N\right),
\ee

\ni if $\tilde{\lambda} = 
-(\sigma - 1)v/2 + \lambda$. 
By Sylvester's theorem, see (\ref{corrtoSyl}), 
taking $X = -\tilde{\lambda}\,I_N$ and $u = v = \sqrt{w}\,1_N$,
we find:

\be
\det\left(
\,w\,1_N\otimes 1_N - \tilde{\lambda}\,\right) \= (-\tilde{\lambda})^{N-1}\,\left(\,
w\,N - \tilde{\lambda}\,\right).
\ee

Thus the latter matrix has eigenvalues $w\,N$ once and 0 repeated $N-1$ times; hence
$A$ has eigenvalues $-(\sigma - 1)v/2 + w\,N$ and $-(\sigma - 1)v/2$ repeated $N-1$ times.
By assumption (\ref{reqtwo}), $A$ is negative definite. 

Next, diagonalize $A$ as :

\be
O\,A\,O^{-1} \= \diag\,(- a_k),
\ee

\ni where $O$ is an orthogonal matrix and $a_k > 0$. Make the change of variables:

\be
z_k \= \sqrt{a_k}\,\sjN\,O_{k,j}\,\left(\,x_j - b_j\,\right).
\ee

\ni Writing $\tilde{\psi}(z_1,...) = \psi(x_1,...)$, the inequality to be proven
becomes:

\bar
\no && \int\,\prod\,dz_k\,\tilde{\psi}^2(z_1,...)\,\left\{\,- \skN\,z^2_k + B\,\right\} \leq\\
\no &&
\left(\,\frac{\sigma - 1}{2\,m}\,\right)\,\int\,\prod\,dz_k\,\skN\,a_k\,\left[\,\frac{\partial
\,\tilde{\psi}}{\partial\,z_k}\,\right]^2.\\
&&\label{ineqtoprove}
\ear

Writing $\delta = (\sigma - 1)\,v/2 - w\,N$, we have $a_k \geq \delta$ for all `$k$';
thus it suffices to prove:

\bar
\no && \int\,\prod\,dz_k\,\tilde{\psi}^2(z_1,...)\,\left\{\,- \skN\,z^2_k + B\,\right\} \leq\\
\no &&
\left(\,\frac{(\sigma - 1)\,\delta}{2\,m}\,\right)\,\int\,\prod\,dz_k\,\skN\,\left[\,\frac{\partial
\,\tilde{\psi}}{\partial\,z_k}\,\right]^2.\\
&&\label{ineqtoprovea}
\ear

Here we can profit from recalling the Dispersion Lemma\footnote{Commonly known as the
``Uncertainty Principle", reflecting Heisenberg 
in his 1927 paper claiming that it expresses
something important about measuring a particle's position or momentum. (Momentum enters
by expressing the second factor {\em via} the Fourier transform.) We \Schists, rejecting
a statistical interpretation of the wavefunction, regard it
as a simple lemma in Fourier Analysis and nothing more.}
 in $N$ dimensions:

\be
\ipdzk\,\phi^2(z_1,...)\skN\,z_k^2\,\cdot\,\ipdzk\,\skN\,
\left[\,\frac{\partial
\,\phi}{\partial\,z_k}\,\right]^2 \Geq \frac{N^2\,||\phi||^4_2}{4},
\ee

\ni valid for $\phi \in L^1(R^N) + L^2(R^N)$.\footnote{See \cite{FandS}, p. 209 second paragraph,
and p. 213, Corollary 2.8.}

Let:

\be
y = \ipdzk\,\tilde{\psi}^2(z_1,...)\skN\,z_k^2,
\ee

\ni and, noting that, due to the Jacobian factor when changing variables,

\be
\ipdzk\,\tilde{\psi}^2 \= \prod\,\sqrt{a_k},
\ee

\ni it suffices to prove:

\be
- y + \tilde{B} \Leq \frac{C}{y},
\ee

\ni where 

\bar
\no \tilde{B} &\=& B\,\prod\,\sqrt{a_k};\\
\no C &\=&
\frac{(\sigma - 1)\,\delta\,N^2\,\prod\,a_k}{8\,m}.\\
&&
\ear

We have a quadratic:

\be
y^2 - \tilde{B}\,y + C \geq 0;
\ee

\ni and so it suffices to know:

\be
\tilde{B}^2 - 4\,C < 0.\label{disc}
\ee

For $B$ we find from (\ref{ABeqns}):

\be
B = \left(\,w\,S\,\right)^2\,1_N^t\,A^{-1}\,1_N.
\ee

\ni Writing

\be
\alpha \= 1_N^t\,A^{-1}\,1_N,
\ee

\ni and
plugging everything into (\ref{disc}) yields:

\be
w < \left\{\,\frac{(\sigma-1)\,\delta\,N^2}{2\,m\,\alpha^2}\,\right\}^{1/4}
\left[\,\frac{1}{|S|}\,\right].\label{plugging}
\ee

Note that $\delta < a_k < (\sigma - 1)\,v/2$.
Hence $A^{-1} < \delta^{-1}$ and:

\be
 \alpha \= 1_N^t\,A^{-1}\,1_N \Leq \delta^{-1}\,1_N^t\,1_N \= \delta^{-1}\,N.
\ee
Thus it suffices to know:

\be
w < \left\{\,\frac{(\sigma-1)\,\delta^3\,}{2\,m}\,\right\}^{1/4}
\left[\,\frac{1}{|S|}\,\right].\label{sufftwo}
\ee

\ni which yields (\ref{reqtwo}).

Proof of (\ref{fifthcondition}):

Define an operator on $\LtRN$ by:
 
\be
\Omega \= \ovx\,\left[\,\ootri + V\,\right]^{-1}
\,\ovx.
\ee

We need to learn whether $\Omega$ is a bounded operator, and how large
$<\psi\,\Omega\,\psi>$ can be. For this task, the matrix coefficients:

\be
<\psi_{\un}\,|\,\ovx|\,\psi_{\um}>
\ee

\ni will prove useful. Here we can exploit some well-known identities concerning the
Hermite functions. For the reader's edification here are the first few:

\be
H_0(x) \= 1;
\ee

\be
H_1(x) \= 2x;
\ee

\be
H_2(x) \= 4\,x^2 - 2;
\ee

\be
H_3(x) \= 8\,x^3 - 12\,x.
\ee

These polynomials can be computed from the recurrence formula:

\be
H_{n+1} \= 2\,x\,H_n - 2\,n\,H_{n-1}.
\ee

\ni which can also be written:

\be
x\,H_n \= \half\,H_{n+1} + n \, H_{n-1}.\label{Hform}
\ee

Writing the one-dimensional eigenfunction as:
\be
\psi_n(x) \= c_n\,
\exp\left\{\, - m\,\omega\,x^2/2\,\right\}\,H_n(x),
\ee

\ni we can use (\ref{Hform}) to obtain:

\be
<\psi_k\,|\,x\,|\,\psi_n> \= <\psi_k\,|\,\left(\,\frac{c_n}{2\,c_{n+1}}\,\right)\,\psi_{n+1}
\+ \left(\,\frac{n\,c_n}{2\,c_{n-1}}\,\right)\,\psi_{n-1}>,
\ee

\ni from which we conclude that 

\be
<\psi_k\,|\,x\,|\,\psi_n> \= 
\begin{cases}
\frac{c_n}{2\,c_{n+1}}, \ph \hbox{if} \ph k = n + 1;\\
\frac{n\,c_n}{2\,c_{n-1}}, \ph \hbox{if} \ph k = n - 1;\\
0, \ph \hbox{otherwise}.
\end{cases}
\ee

We have

\be
c_n \= \frac{c}{\sqrt{2^n\,n!}},
\ee

\ni where $c$ is a constant; hence we get

\be
<\psi_k\,|\,x\,|\,\psi_n> \= 
\begin{cases}
\sqrt{\frac{n+1}{2}}, \ph \hbox{if} \ph k = n + 1;\\
\sqrt{\frac{n}{2}}, \ph \hbox{if} \ph k = n - 1;\\
0, \ph \hbox{otherwise}.
\end{cases}
\ee

Using the orthoginalty relations for the eigenfunctions we have: 

\be
<\psi_{\un}\,|\,\ovx|\,\psi_{\um}> \= \sjN\,<\psi_{n(j)}\,|\,x_j\,|\,\psi_{m(j)}>.
\ee

Returning to the operator $\Omega$ we now have for its matrix elements the development:

\be
<\psi_{\un}\,|\,\Omega\,|\,\psi_{\um}> \= \sum_{\um'}\,<\psi_{\un}\,|\,\ovx\,|\,\psi_{\um'}>
\,\cE_{\um'}^{-1}\,<\psi_{\um'}\,|\,\ovx\,|\,\psi_{\um}>.
\ee

In order to state some general result define:

\be
Q_{\un,\um} \= <\psi_{\un}\,|\,\ovx|\,\psi_{\um}> 
\ee

\ni and

\be
R_{\un,\um} \= <\psi_{\un}\,|\,\Omega\,|\,\psi_{\um}>;
\ee 

\ni then using (\ref{eigenvl}) we can write the matrix elements of $\Omega$ as:

\be
R_{\un,\um} \= (\,\omega\,)^{-1}\,\sum_{\um'}\,Q_{\un,\um'}\,
\left\{\,||\um'|| + N/2\,\right\}^{-1}\,Q_{\um',\um},
\ee

\ni where $||\um|| = \skN\,m(k).$

From previous results we see that $Q_{\un,\um} \neq 0$ only if, for some $j$,
$1 \leq j \leq N$, $|n(j) - m(j)| = 1$ and $n(k) = m(k)$ for all $k \neq j$,
in which case

\be
Q_{\un,\um} \= \sqrt{\frac{\max\{n(j),m(j)\}}{2}}.
\ee

From this there follows that $R_{\un,\um} \neq 0$ only if $\un$ and $\um$ differ at most
at two entries and we have the bound:

\be
R_{\un,\um} \leq \frac{\sqrt{m'(j) + 1}\,\sqrt{m'(k) + 1}}{||\um'|| + N/2},
\ee

\ni for some $\um'$, $j$, and $k$. Using the inequality: $a\,b\leq 1/2(a^2 + b^2)$
then yields:

\bar
\no R_{\un,\um} &\leq& \frac{\sqrt{m'(j) + 1}\,\sqrt{m'(k) + 1}}{||\um'|| + N/2}\\
\no &\leq& \frac{(1/2)\,\left\{\,m'(j) + m'(k) + 2\right\}}{||\um'|| + N/2}\\
\no &\leq& \frac{\,||\um'|| + 1}{||\um'|| + N/2}\\
\no &\leq& 2.\\
&&
\ear

We are now prepared to apply the following theorem about matrix operators. Let
$\cC$ be a countable set and $\cR = \Runum$ a matrix of real numbers defined on $\cC\times\cC$.
We regard $\cR$ as an operator on $l^2(\cC)$, the square-integrable sequences
on $\cC$, which we denote $\cala = \{\,\aun\,\}$:

\be
||\cala||^2 \= \sunC\,\aun^2 < \infty.
\ee

{\em via} the (formal) sum

\be
\left(\,\cR\,\cala\,\right)_{\un} \= \sumC\,\Runum\,\aum.
\ee

\begin{quote}
{\bf Theorem 4} Suppose:

(a) $\Runum = R_{\um,\un}$ and $|\Runum| \leq R < \infty$ for all $\un$ and $\um$ in $\cC$;

(b) For every $\un \in \cC$ there is a set $\Lun \subset \cC$ such that $\Runum = 0$
except if $\um \in \Lun$;

(c) $|\Lun| \leq \Lambda < \infty$ for all $\un \in \cC$.

Then $\cR$ defines an operator on $l^2(\cC)$ which is moreover bounded:

\be
\sunC\,\left\{\,\sumC\,\Runum\,\aum\,\right\}^2 \leq R^2\,\Lambda^2\,\sunC\,\aun^2.
\ee

\ni which can be written as the operator bound:

\be
||\cR||_{\hbox{op}} \leq R\,\Lambda.
\ee

\end{quote}

{\bf Proof of Theorem 4}:

\bar
\no && \sunC\,\left\{\,\sumC\,\Runum\,\aum\,\right\}^2 \= \\
\no && \sunC\,\left\{\,\sumC\,1[\,\um \in \Lambda_{\un}\,]\,\Runum\,\aum\,\right\}^2 \leq \\
\no && \sunC\,\left\{\,\sumC\,1[\,\um \in \Lambda_{\un}\,]\,\Runum^2\,\right\} \, \times\\
\no && \left\{\,\sumC\,1[\,\um \in \Lambda_{\un}\,]\,\aum^2\,\right\} \=  \\
\no && R^2\,\Lambda\,\sumC\,\aum^2\,\left\{\,\sunC\,1[\,\um \in \Lambda_{\un}\,]\,\right\} \=  \\
\no && R^2\,\Lambda^2\,\sumC\,\aum^2.\\
&&
\ear

\begin{quote}

{\bf Theorem 5}. 

(a) The operator $\Omega$ on $\LtRN$ is bounded and in fact:

\be
||\Omega|| \leq 8\,N^2.
\ee

(b) If $N \geq 2$ there exists a real-valued $\psi \in \LtRN$ such that

\be
<\psi\,|\,\Omega\,|\,\psi> \ph > \ph N\,(N+1)\,||\psi||^2.
\ee

\end{quote}
 
{\bf Proof of the Theorem}: We identify $\LtRN$ with $l^2(\cC)$ using the eigenfunctions 
of the linear model, as discussed above. We first note that

\be
N\,(N+1) \leq |\Lambda_{\un}| \leq 4\,N^2.
\ee

\ni This follows from the above remark about when $\Runum$ can be nonzero.
The upper bound refers to the case with e.g., $n(k) > 0$ for all $k$,
in which case there are $N^2$ choices of at most two entries to change, and 
two choices ($\pm 1$) for each. The lower bound comes from the ground state: $n(k) = 0$
for all $k$. 

To prove (a), we use the above calculations plus the remark that $R = 2$ and $\Lambda < 4\,N^2$.

To prove (b), let there be a sequence $\{a_{\un}\}$ having the properties:

\bar
\no a_{\un} &\geq& 0 \ph \hbox{for all } \ph \un \in \cC;\\
\no a_{\um} &\geq& c\,a_{\un} \ph \hbox{if} \ph m \in \Lambda_{\un},\\
&&
\ear

\ni for some $c > 0$. Then we have:

\bar
\no <\psi\,|\,\Omega\,|\,\psi> &\=& \sum_{\un \in \cC}\,a_{\un}\,\sum_{\um \in \cC} 
1[\um \in \Lambda_{\un}]\,\Runum\,a_{\um}\\
\no &\geq& c\,R\,\sum_{\un \in \cC}\,a_{\un}^2\,|\Lambda_{\un}|\\
\no &\geq& c\,R\,N\,(N+1)\,\sum_{\un \in \cC}\,a_{\un}^2.\\
&&
\ear

An example of a sequence $\{a_{\un}\}$ is:

\be
a_{\un} \= \left(\,\frac{1}{n(1) + N}\,\right)\,
 \left(\,\frac{1}{n(2) + N}\,\right)\,\dots\,
\left(\,\frac{1}{n(N) + N}\,\right).
\ee

Then for a sequence $\{a_{\um}\} \in \Lambda_{\un}$ we get:

\bar
\no a_{\um} &\=& \left(\,\frac{1}{n(1) + N}\,\right)\,\dots\,
 \left(\,\frac{1}{n(j) + N \pm 1}\,\right)\,\dots\,
\left(\,\frac{1}{n(k) + N \pm 1}\,\right)\,\dots\\
\no &\=&  
\left(\,\frac{n(j) + N}{n(j)+ N \pm 1}\,\right)\,
\left(\,\frac{n(k) + N}{n(k) + N \pm 1}\,\right)\,a_{\un},\\
&&
\ear

\ni and the two ratios in the last line are bounded below by 2/3 for $N \geq 2$.
So we can take $c = 0.67$ for this example.

Now to the Proposition: there are several scenarios in which we can find a wavefunction $\psi$
satisfying conditions (i) - (v).

Scenario A: $S = 0$; $\psi$ is real; and

\be
\frac{\sigma}{4\,<\psi\,|\Omega\,|\,\psi>} < \frac{(\sigma -1)\,v}{2\,N},\label{ineqx}
\ee

\ni for some $\sigma > 1$. For the first condition, $S$ will be zero if the wavefunction
is symmetric or antisymmetric under reflection in the origin; i.e.,

\be
\psi(-x_1,-x_2,...,-x_N) \= \pm\,\psi(x_1,x_2,...,x_N).
\ee

We can construct such wavefunctions from the eigenfunctions of the linear model, as follows.
Note that the eigenfunction $\psi_{\un}$ will contain the product
of Hermite polynomials:

\be 
\prod_{k=1}^N\,H_{n(k)}(\sqrt{m\,\omega}\,x_k).
\ee

\ni The Hermite polynomial $H_n(x)$ contains all even/odd powers of $x$ if $n$ is even/odd.
Therefore

\be 
\prod_{k=1}^N\,H_{n(k)}(- \sqrt{m\,\omega}\,x_k) \= (-1)^{\# \hbox{odd $n(k)$}}
\prod_{k=1}^N\,H_{n(k)}(\sqrt{m\,\omega}\,x_k).
\ee

Since

\be
n(k) \= 
\begin{cases} 
2r(k),& \hbox{if}\ph n(k)\ph \hbox{is even};\\
2r(k)+1,& \hbox{if}\ph n(k)\ph \hbox{is odd};
\end{cases}
\ee

\ni for some integers $r(k)$, clearly:

\be 
(-1)^{\# \hbox{odd $n(k)$}} \= (-1)^{||\un||}.
\ee

\ni Thus $\psi_{\un}$ is symmetric/antisymmetric under reflection 
depending on whether $||\un|| = \sum n(k)$
is even/odd.

So to construct symmetric or antisymmetric wavefunctions, we simply restrict the
allowed sequences $\un$ to have even or odd sums. Let $\cCe$ and $\cCo$ denote
these subsets respectively. Then for, e.g., a symmetric wavefunction we pose

\be
\psi \= \sum_{\un \in \cCe}\,a_{\un}\,\psi_{\un}.
\ee

For the proof of Theorem 4, I produced a wavefunction such that the quantity appearing in the
denominator of the left side of (\ref{ineqx}) is O$(N^2)$. This construction can be imitated
restricting the sequences to lie in $\cCe$ or $\cCo$. Note that given e.g., $\un \in \cCe$,
an allowed $\um \in \Lambda_{\un}$ will differ from $\un$ by adding $\pm 1$ at most at
two entries; thus $\um \in \cCe$ as well. The bounds follow as before.

Scenario B: 

Both (\ref{reqtwo}) (even if $S \neq 0$) and (\ref{reqthree}) hold.

\section*{Math Appendix 5: Numerical methods}

The Hamiltonian model (meaning the RDS) was simulated using the Tao method, \cite{Tao}.
I usually used 20,000 time steps and checked that the routine preserved both the energy and
the norm to 5 decimal places.

The RDS+JDS system was simulated by making Tao updates for the first half 
and ``RK2" for the second half, see \cite{NumRC}, p. 569-70. 

Determinants were computed using the LU decomposition, see
\cite{NumRC}, p. 46. 

The routines were written in the `C' language and run on a 20-year-old HP table-top computer.


\begin{thebibliography}{9}

\bibitem{paperI}
Wick, W. D. ``On Non-Linear Quantum Mechanics and the Measurement Problem I: Blocking Cats", 
ArXiv 1710.03278. Published October 2017.

\bibitem{paperII}
Wick, W. D. ``On Non-Linear Quantum Mechanics and the Measurement Problem  
II: The Random Part of the Wavefunction", ArXiv 1710.03800, published October 2017.  

\bibitem{paperIII}
Wick, W. D. ``On Non-Linear Quantum Mechanics and the Measurement Problem  
III: Poincar{\' e} Probability and ... Chaos?" ArXiv 1803.1126v1 published March 2018.

\bibitem{Skokos}
Skokos, Charalampos. ``The Lyapunov Characteristic Exponents and their computation."
ArXiv 0811.0882v2. Published January 2009.

\bibitem{MandH}
Murray, N. and Holman, M. ``The Origin of Chaos in the Outer Solar System."
{\em Science}, 19 March 1999. 283: 1877-81.

\bibitem{Wick2013}
Wick, W.D. {\em Fitting Non-Linear Stochastic Models to Data in Biology and Medicine:
Philosophy and technique}. Available from Amazon (e-book or print version). (2013).

\bibitem{Bill}
Faris, W. G. {\em Self-Adjoint Operators}. Lecture Notes in Mathematics 433.
Springer, NY. (1975).

\bibitem{FandS}
Folland, G.B. and Sitaram, A. ``The Uncertainty Principle: A Mathematical Survey".
J. Fourier Anal. and App. V3,no. 3, p. 208-238. (1997).

\bibitem{NumRC}
Press, W. H. {\em et al}. {\em Numerical Recipes in C: The Art of Scientific Computing}.
Cambridge U Press. Printed in the US, NY, NY. (1989).

\bibitem{Tao}
Tao, M. ``Explicit symplectic approximation of nonseparable Hamiltonians: 
algorithm and long-time performance." Arxiv 1609.02212v1, September 2016; 
Phys Rev E. 94: 043303 2016.

\bibitem{Hall}
Halln{\'a}s, M. {\em Exactly solved many-body systems in one dimension}. 
Thesis, Department of Physics, Royal Institute of Technology, Stockholm, Sweden. (2005).


\end{thebibliography}
\end{document}